\documentclass[
    ,final            
  ]
  {aipproc}

\layoutstyle{6x9}

\newcommand{\bra}[1]{\langle#1|}                  
\newcommand{\ket}[1]{|#1\rangle}                  
\newcommand{\rmeh}{{\rm e}}
\newcommand{\rmih}{{\rm i}}

\newcommand{\be}{\begin{equation}}
\newcommand{\ee}{\end{equation}}
\newcommand{\bea}{\begin{eqnarray}}
\newcommand{\eea}{\end{eqnarray}}
\newcommand{\beas}{\begin{eqnarray*}}
\newcommand{\eeas}{\end{eqnarray*}}

\newcommand{\ham}{\hat{H}}                          
\newcommand{\dm}{\hat{\rho}}                        
\newcommand{\imag}{{\rm i}}                        
\newcommand{\impart}{{\rm Im}}                     
\newcommand{\repart}{{\rm Re}}                     
\newcommand{\BEA}{\begin{eqnarray}}
\newcommand{\EEA}{\end{eqnarray}}

\newcommand{\ini}{\ensuremath{\ket{\uparrow \dots \uparrow \downarrow
      \dots \downarrow}}}
\newcommand{\ketini}{\ensuremath{\ket{\mathrm{ini}}}}
\newcommand{\init}{\ensuremath{\mathrm{ini}}}

\newcommand{\Fig}[1]{Fig.~\ref{#1}}
\newcommand{\pdag}{{\phantom{\dagger}}}

\begin{document}

\title{Methods for Time Dependence in DMRG}


\author{Ulrich Schollw\"ock}{
  address={Institute for Theoretical Physics C\\
  RWTH Aachen University\\
  D-52056 Aachen, Germany}
}

\author{Steven R. White}{
  address={Department of Physics and Astronomy\\
University of California at Irvine\\
Irvine, California 92697
}
}

\begin{abstract}
A major advance in density-matrix renormalization group (DMRG)
calculations has been achieved by the invention of highly efficient
DMRG techniques for the simulation of real-time dynamics of strongly
correlated quantum systems in one dimension. Starting from established
linear-response techniques in DMRG and early attempts at real-time
dynamics, we go on to review two current methods which both implement 
the idea of adapting the effective Hilbert space of DMRG to the
quantum state evolving in time. We also give an outlook on extensions to finite
temperature calculations.
\end{abstract}
\maketitle


\section{Introduction}

The physics of strongly correlated quantum systems continues to pose
major challenges in experimental and theoretical physics. While both
experiment and theory have focused on static, thermodynamic or at most
linear-response quantities in the past, recently questions which explicitly
involve the out-of-equilibrium time-dependence of such quantum systems
have come to the foreground. 
These questions arise in the context of transport
far from equilibrium or of decoherence, particularly as the size
of devices continues to shrink towards the atomic scale. However,
perhaps the most striking example for
this is provided by the progress in preparing dilute ultracold bosonic and
also fermionic alkali gases. Subjected to an optical lattice, these
gases are arguably the purest realization of the typical model
Hamiltonians of strong correlation physics, such as the Hubbard 
model\cite{Grei02,Kohl05}.
More importantly, the interaction parameters can be tuned
experimentally on quantum mechanically relevant time-scales over a huge
range, while being known precisely from microscopic calculations. 
From a theoretician's point of view, this situation is almost ideal, and has
stimulated great interest in the development of time-dependent methods. 

For linear response, exact diagonalization can provide
detailed results for small systems, and quantum Monte Carlo can provide
coarse resolution results after analytic continuation from imaginary time,
for systems without the sign problem. 
Outside the linear response regime, almost the only tool available has
been the diagonalization of very small clusters.

In this review, the emphasis is on recent extensions of the
density-matrix renormalization group method (DMRG) \cite{Whit92,Whit93,Scho04} 
into the real-time domain which make it the currently most powerful method for 
such problems. Following up on early attempts to extend DMRG to real-time, 
input from quantum information theory has led to the formulation of
two DMRG algorithms for real-time evolutions. We set out with a
reminder about the basic ideas of DMRG, review linear response
calculations with DMRG and move on to early attempts in the time domain. 
We then explain how the TEBD algorithm of Vidal \cite{Vida04} beautifully reflects
fundamental structures of DMRG and hence can be easily used to extend 
DMRG to the time-domain \cite{Dale04,Whit04}. However, this approach has shortcomings for
longer-ranged interactions, which can be circumvented in yet another
modification of DMRG at some cost in efficiency \cite{Whit04c}. 
The range and power of
both methods are discussed based on ``real-life'' applications.

\section{Basic Ideas of DMRG}

Several good descriptions of the DMRG algorithms exist in the literature
\cite{Whit92,Whit93,Scho04}. Rather than repeat these descriptions, here we
summarize the most important ideas of DMRG.

The first key idea is the description of a collection of sites, or block,
in terms of a limited set of basis states and operator matrices
between those states. These states and matrices are defined by a set
of basis transformations as sites are successively added to the block.
This representation is due to Wilson and is a key feature of his
numerical renormalization group (NRG)\cite{Wils75}. Let the block at the
beginning of step $\ell$ be described by a set of states $\{\ket{i}\}$. In
this step site $\ell$ (states $\{ \ket{\sigma} \}$) is added to the block. 
The new states describing the larger block $\{ \ket{i'} \}$ are given by
\[
\ket{i'} = \sum_{\sigma,i} A^{\ell}_{ii'}[\sigma]
\ket{i}\otimes\ket{\sigma} .
\]
The number of states is kept approximately constant at $m$, so the set of states
$\{ \ket{i'} \}$ is incomplete. If the states $\{ \ket{i'} \}$ were described in
detail in terms of the sites, the computational effort would grow
exponentially despite the truncation to a constant number of states.
Instead, the $m\times m$ matrices for the Hamiltonian and other
operators give all the detail needed to construct the Hamiltonian 
at larger length scales. These matrices are transformed to the new
basis at each step using the transformation matrices $A$. In this
way, the computational effort remains constant as $O(m^3)$.

The second key idea of DMRG is to choose the states to keep as eigenstates 
of the reduced density matrix (RDM) of the block. In Wilson's NRG
approach, one kept the lowest energy eigenstates of the block
Hamiltonian. This choice works for the special Hamiltonians devised by
Wilson for impurity systems, but fails for more general lattice
systems. Using the density matrix eigenstates can be shown to be
optimal for reproducing the wavefunction as well as the RDM. Let the
block have states $\ket{i}$ and the rest of the system, referred to as the
environment, have states $\ket{j}$. Then the wavefunction of the whole
system is written as
\[
\ket{\psi} = \sum_{ij} \psi_{ij} \ket{i} \otimes \ket{j}
\]
and the coefficients of the RDM $\rho$ are
\[
\rho_{ii'} = \sum_j \psi_{ij} \psi^*_{i'j} .
\]
The eigenvalues of $\rho$ are the probabilities of the block being in
the corresponding eigenstate, and if the probability is neglible, the
eigenstate can be omitted from the basis. The RDM can also be built
by summing over several states $\ket{\psi^a}$, with arbitrary weights,
representing the probability of each $\psi^a$ in a mixed state of the
system. In this case each $\ket{\psi^a}$ is said to be {\it targetted}, an
important concept for time-dependent DMRG.

The RDM depends on the enviroment through $\ket{\psi}$. In DMRG, both the
block and the environment are described approximately using basis
sets of size $\sim m$. This leads to the third key idea of DMRG, the
idea of sweeping back and forth to produce a self-consistent, accurate
representation for both parts of the system. In Fig. 1 we show the
most common superblock configuration in DMRG. In the sweeping
procedure the dividing line between the left and right block moves
back and forth between the ends of the system. The block which is
growing is treated as the system block; the other is the enviroment,
with the roles reversed when the direction is reversed. The system
block in each case obtains an improved basis during the sweep. In
Wilson's NRG, there was no sweeping and no feedback from the low
energy scales to the high energy scales. For a general lattice system,
not divided by energy scales, feedback is necessary.

\begin{figure}
  \includegraphics[height=.05\textheight]{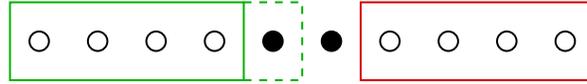}
  \caption{The standard DMRG superblock configuration, in which the left
central site is being added onto the left block.}
\end{figure}

If we trace back through the basis set transformations which led to
the basis of a block, one obtains an explicit representation of the
states 
\[
\ket{i_\ell} = \sum_{\sigma_1\ldots \sigma_\ell} 
[A^{1}[\sigma_1] A^{2}[\sigma_2]\ldots A^{\ell}[\sigma_\ell]]_{i_\ell}
\ket{\sigma_1 \ldots \sigma_\ell}
\]
where $A^1$ is a vector for each $\sigma_1$ and the rest of the $A$'s are
matrices. We can write the basis states for the right block similarly.
Alternatively, we can let the dividing line be all the way on the
right end of the system, in which case we can write the following
{\it matrix product}  expression for $\ket{\psi}$:
\[
\ket{\psi} = \sum_{\sigma_1\ldots \sigma_L} 
A^{1}[\sigma_1] A^{2}[\sigma_2]\ldots A^{L}[\sigma_L] \ket{\sigma_1 \ldots 
\sigma_L}
\]
where $A^{L}[\sigma_L]$ is a vector for each value of $\sigma_L$, so that the
product of the $A$'s is a scalar. The matrix product representation
for $\ket{\psi}$ was first developed in a DMRG context by Ostlund and
Rommer\cite{Ostl95}, but its usefulness was not widely appreciated at
the time. More recently, the matrix product representation has become
very important as a route to improve the capabilities of and to
generalize DMRG \cite{Vers04}.

\begin{figure}
  \includegraphics[height=.2\textheight]{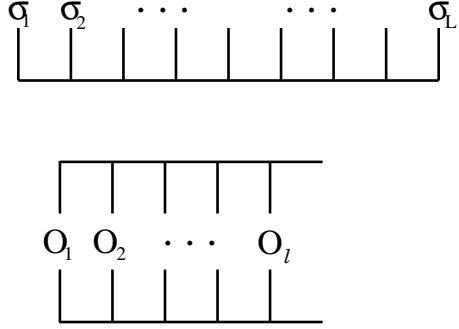}
  \caption{Diagrams for the matrix product representation of the wavefunction
and an operator on a block.}
\end{figure}

A useful diagrammatic form of the matrix product
representation \cite{Vers04} is
illustrated in Fig. 2. The upper diagram represents the wavefunction. 
Here each intersection of lines is associated with a
site and represents a matrix (here the $A^{\ell}[\sigma_\ell]$), or
more generally, a tensor. The vectors $A^1$ and $A^L$ are reprecented
by right-angle segments.  Interior line segments represent indices
which are summed over, whereas the ends of segments sticking out of
the figure represent external indices labeling states. As another example,
let an operator
$\hat O$ be defined as a product of site operators on the left block, 
$\hat O_1 \ldots \hat O_\ell$, where many of the $\hat O_i$ may be identity
operators. The lower diagram represents the matrix for this operator in the basis
of the left block. Each of the two lines sticking out on the right represents
indices running over the $m$ states of the left block.

\section{DMRG dynamics}
While the original DMRG algorithm seems to be limited to the
calculation of equilibrium properties such as ground state
correlations, it can in fact be extended to linear response
quantities. For some operator $\hat{A}$, we define a (time-dependent) Green's 
function at $T=0$ in the Heisenberg picture by
\begin{equation}
    \imag G_{A}(t'-t) = \bra{0} \hat{A}^\dagger(t') \hat{A}(t) \ket{0}  
    \label{eq:deftimecorrel}
\end{equation}
with $t' \geq t$ for a time-independent Hamiltonian $\ham$. Going to
frequency-space, the Green's function reads
\begin{equation}
    G_{A}(\omega+\imag\eta)= \bra{0} \hat{A}^\dagger \frac{1}{E_{0}+\omega+\imag\eta-\ham} 
    \hat{A} \ket{0} ,
    \label{eq:Greensfunction}
\end{equation}
where $\eta$ is some positive number to be taken to 
zero at the end. We may also use the spectral or Lehmann 
representation of correlations in the eigenbasis of $\ham$,
\begin{equation}
    C_{A}(\omega) = \sum_{n} |\bra{n} \hat{A} \ket{0}|^2 
    \delta(\omega+E_{0}-E_{n}).
    \label{eq:Lehmann}
\end{equation}
$C_{A}(\omega)$ is related to $G_{A}(\omega+\imag\eta)$ as
\begin{equation}
    C_{A}(\omega) = \lim_{\eta\rightarrow 0^+} - \frac{1}{\pi} \impart G_{A}(\omega+\imag\eta) .
    \label{eq:etalimit}
\end{equation}    
The role of $\eta$ in DMRG calculations is threefold:
First, it ensures causality in Eq.\ (\ref{eq:Greensfunction}). 
Second, it introduces a 
finite lifetime $\tau\propto 1/\eta$ to excitations. 
Third, $\eta$ provides a Lorentzian broadening of $C_{A}(\omega)$,
\begin{equation}
    C_{A}(\omega+\imag\eta) = \frac{1}{\pi} \int d\omega' C_{A}(\omega') 
    \frac{\eta}{(\omega-\omega')^2 + \eta^2},
    \label{eq:lorentzian}
\end{equation}
which serves either to broaden the numerically obtained discrete spectrum of 
finite systems into some ``thermodynamic limit'' behavior or to 
broaden analytical results for $C_{A}$ for comparison to numerical 
spectra where $\eta>0$.

Most DMRG approaches to dynamical correlations center 
on the evaluation of Eq.\ (\ref{eq:Greensfunction}). The first, which 
we refer to as Lanczos vector dynamics, has been 
pioneered by Hallberg \cite{Hall95}, and calculates highly time-efficient, but comparatively 
rough approximations to dynamical quantities adopting the 
Balseiro-Gagliano method \cite{Gagl87} to DMRG. 
The second class of approaches, 
including both the correction vector method \cite{Rama97,Kuhn99} and 
DDMRG (dynamical DMRG) \cite{Jeck02}, 
is also based on pre-DMRG techniques, but is both much more precise
and numerically much more expensive.

Boundary effects due to DMRG-typical open boundary conditions can 
be treated in various ways; two situations should be distinguished. 
If $\hat{A}$ acts locally, such as in the calculation of an optical 
conductivity, one may exploit  that finite $\eta$ exponentially suppresses 
excitations \cite{Jeck02}. As they travel at some speed $c$ through the system, a 
thermodynamic limit $L\rightarrow\infty$ first, $\eta\rightarrow 0$
second may be taken consistently as a single limit with
$\eta=c/L$. For the calculation of dynamical structure 
functions such as obtained in elastic neutron scattering, $\hat{A}$ is 
a spatially delocalized Fourier transform, and another approach must be 
taken. The open boundaries introduce both
genuine edge effects and a hard cut to the wave functions of excited 
states in real space, leading to a large spread in momentum space. To 
limit bandwidth in momentum space, filtering by modifying
$\hat{A}(x)\rightarrow \hat{A}(x)f(x)$ is necessary. The 
filtering function $f(x)$ should be narrow in 
momentum space and broad in real space, while simultaneously strictly
excluding edge sites. For a detailed discussion of such filters, see
\cite{Kuhn99}.
\subsection{Continued fraction dynamics}
\label{subsec:lanczosdynamics}
The technique of {\em continued fraction dynamics} has first been 
exploited by Gagliano and Balseiro \cite{Gagl87} in the framework of exact ground state 
diagonalization. Obviously, the calculation of Green's functions as in Eq.\ 
(\ref{eq:Greensfunction}) involves the inversion of $\ham$ (or more 
precisely, $E_{0}+\omega+\imag\eta-\ham$), a typically 
very large sparse hermitian matrix. This inversion is carried out in 
two, at least formally, exact steps. First, an iterative basis 
transformation taking $\ham$ to a tridiagonal form is carried out.
Second, this tridiagonal matrix is then inverted, allowing the 
evaluation of Eq.\ (\ref{eq:Greensfunction}).

Let us call the diagonal elements of $\ham$ in the tridiagonal form $a_{n}$ and 
the subdiagonal elements $b_{n}^2$. 
The coefficients $a_{n}$, $b_{n}^2$ are obtained as the 
Schmidt-Gram coeffcients in the generation of a Krylov subspace of
unnormalized states starting from some arbitrary state, which we take 
to be the excited state $\hat{A}\ket{0}$:
\begin{equation}
    \ket{f_{n+1}} = \ham\ket{f_{n}} - a_{n} \ket{f_{n}} - b_{n}^2 
    \ket{f_{n-1}} ,
    \label{eq:Krylovsequence}
\end{equation}
with $\ket{f_{0}} = \hat{A}\ket{0}$, and
\begin{equation}
\    a_{n} = \frac{\bra{f_{n}} \ham\ket{f_{n}}}{\langle f_{n} | 
    f_{n} \rangle}, \quad\quad 
    b_{n}^2 =\frac{\bra{f_{n-1}} \ham\ket{f_{n}}}{\langle f_{n-1} | 
    f_{n-1} \rangle} = \frac{\langle f_{n} |f_{n}\rangle}{\langle f_{n-1} | 
    f_{n-1} \rangle}.
\end{equation}    
The global orthogonality of the states $\ket{f_{n}}$ (at least in 
formal mathematics) and the tridiagonality of the new representation (i.e.\
$\bra{f_{i}} \ham \ket{f_{j}}=0$ for $|i-j|>1$) follow 
by induction. It can then be shown quite 
easily by an expansion of determinants that the inversion of 
$E_{0}+\omega+\imag\eta-\ham$ leads to a continued fraction such that 
the Green's function $G_{A}$ reads
\begin{equation}
    G_{A}(z) = \frac{\bra{0} \hat{A}^\dagger \hat{A} \ket{0}}{
    z-a_{0}-\frac{b_{1}^2}{z-a_{1}-\frac{b_{2}^2}{z-\ldots}}} ,
    \label{eq:continuedfraction}
\end{equation}
where $z=E_{0}+\omega+\imag\eta$. This expression can now be 
evaluated numerically, giving access to dynamical correlations. 
Alternatively, one may also exploit 
that upon normalization of the Lanczos vectors $\ket{f_{n}}$ and accompanying 
rescaling of the $a_{n}$ and $b_{n}^2$, the Hamiltonian is 
iteratively transformed into a tridiagonal form in a new approximate 
orthonormal basis. Transforming 
the basis $\{ \ket{f_{n}} \}$  by a diagonalization of the 
tridiagonal Hamiltonian matrix to the 
approximate energy eigenbasis of $\ham$, $\{ \ket{n} \}$ with 
eigenenergies $E_{n}$, the Green's function can be written within this 
approximation as
\begin{eqnarray}
    & & G_{A}(\omega+\imag\eta) = \\
    & & \sum_{n} \bra{0} \hat{A}^\dagger \ket{n}\bra{n}  
    \frac{1}{E_{0}+\omega+\imag\eta-E_{n}}  \ket{n}\bra{n} 
    \hat{A} \ket{0} \nonumber ,
\end{eqnarray}    
where the sum runs over all approximate eigenstates. The 
dynamical correlation function is then given by
\begin{equation}
    C_{A}(\omega+\imag\eta) = \frac{\eta}{\pi} \sum_{n} \frac{|\bra{n} 
    \hat{A} \ket{0}|^2}{(E_{0}+\omega-E_{n})^2 +\eta^2} ,
\end{equation}
where the matrix elements in the numerator are simply the 
$\ket{f_{0}}$ expansion coefficients of the approximate eigenstates 
$\ket{n}$.

In practice, several limitations occur. The iterative generation of the 
coefficients $a_{n}$, $b_{n}^2$ is equivalent to a Lanczos diagonalization of 
$\ham$ with starting vector $\hat{A}\ket{0}$. Typically, the convergence of the 
lowest eigenvalue of the transformed tridiagonal Hamiltonian to the 
ground state eigenvalue of $\ham$ will have happened after 
$n \sim O(10^2)$ iteration steps for standard model Hamiltonians. 
Lanczos convergence is however accompanied by numerical loss of global 
orthogonality which computationally is ensured only locally,
invalidating the inversion procedure.  With $\hat{A}\ket{0}$ as starting vector, convergence 
will be fast if $\hat{A}\ket{0}$ is a long-lived excitation (close to an eigenstate) 
such as would be the 
case if the excitation is part of an excitation band; this will 
typically not be the case if it is part of an excitation continuuum.
Moreover, $\ham$ itself is not exact, and its repeated application
generates further errors.

\begin{figure}
    \includegraphics[scale=0.32,angle=270]{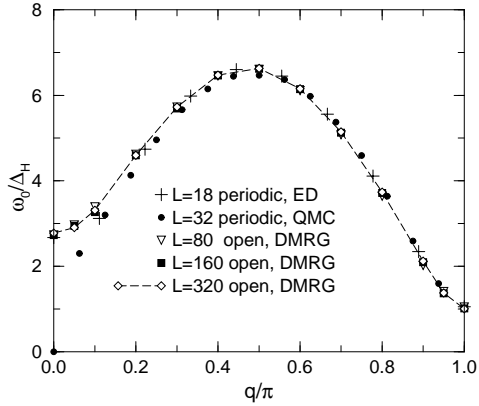}
\caption{Single magnon line of the $S=1$ Heisenberg AFM from exact 
diagonalization, quantum Monte Carlo and DMRG for various system 
sizes and boundary conditions. From \cite{Kuhn99}.}
\label{fig:singlemagnon_spin1}
\end{figure}

As an example for the excellent performance of this method, one may 
consider the isotropic spin-1 Heisenberg chain, where the single 
magnon line is shown in Fig.\ \ref{fig:singlemagnon_spin1}. Exact 
diagonalization, quantum Monte Carlo and DMRG are in excellent 
agreement, with the exception of the region $q\rightarrow 0$, where 
the single-magnon band has disappeared into a two magnon continuum. 
Here Lanczos vector dynamics does a poor job reproducing the peak of the
spectral function just above the bottom of the continuum, which has a gap of
twice the Haldane gap $\Delta_H$ at $q=0$.
	
\begin{figure}
    \includegraphics[scale=0.32,angle=270]{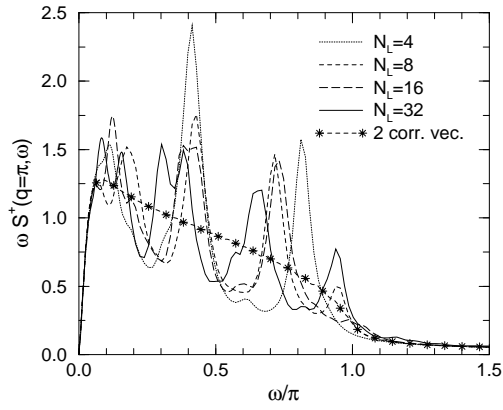}
\caption{Spectral weight $S^+(q=\pi,\omega)$ 
of the $S=1/2$ Heisenberg AFM from Lanczos 
vector and correction vector DMRG. $N_{L}$ indicates the number of 
target states; $M=256$. Note that spectral weight times $\omega$ is 
shown. From \cite{Kuhn99}.}
\label{fig:singlemagnon_spin12}
\end{figure}

The intuition that excitation continua are badly approximated by 
a sum over some $O(10^2)$ effective excited states is further 
corroborated by considering the spectral weight function $S^+(q=\pi,\omega)$ 
[use $A=S^+$ in Eq.\ (\ref{eq:etalimit})] for a 
spin-1/2 Heisenberg antiferromagnet. As shown in Fig.\ 
\ref{fig:singlemagnon_spin12}, Lanczos vector dynamics roughly 
catches the right spectral weight, including the $1/\omega$ 
divergence, as can be seen from the essentially exact correction 
vector curve, but no convergent behavior can be observed 
upon an increase of the number of targeted vectors. The very fast 
Lanczos vector method is thus certainly useful to get a quick 
overview of spectra, but not suited to detailed quantitative 
calculations of excitation continua, only excitation bands. 

\subsection{Correction vector dynamics}
\label{subsec:correctiondynamics}
Even before the advent of DMRG, another way to obtaining more precise 
spectral functions had been proposed in \cite{Soos89}; it was 
first applied using DMRG in \cite{Rama97} and \cite{Kuhn99}. After 
preselection of a fixed frequency $\omega$ one may introduce a 
{\em correction vector} 
\begin{equation}
    \ket{c(\omega+\imag\eta)} = \frac{1}{E_{0}+\omega+\imag\eta-\ham}\hat{A} \ket{0} ,
    \label{eq:correctionvector}
\end{equation}
which, if known, allows for trivial calculation of the Green's 
function and hence the spectral function at this particular frequency:
\begin{equation}
    G_{A}(\omega+\imag\eta)=\langle A \ket{c(\omega+\imag\eta)} .
    \label{eq:Greensfunctionfromcorrection}
\end{equation}    
The correction vector itself is obtained by solving the large sparse linear 
equation system given by
\begin{equation}
(E_{0}+\omega+\imag\eta-\ham)\ket{c(\omega+\imag\eta)} =   \hat{A} \ket{0} .
\label{eq:sparseequation}
\end{equation}
To actually solve this nonhermitean equation system, the current 
procedure is to split the correction vector into real and imaginary 
part, to solve the hermitean equation for the imaginary part and 
exploit the relationship to the real part: 
\begin{eqnarray}
    & & [(E_{0}+\omega-\ham)^2+\eta^2] \impart \ket{c(\omega+\imag\eta)} =  -\eta 
    \hat{A} \ket{0} \\
    & & \repart \ket{c(\omega+\imag\eta)} = \frac{\ham-E_{0}-\omega}{\eta} 
    \impart \ket{c(\omega+\imag\eta)}
    \label{eq:correctionvectorsystem}
\end{eqnarray}    
The standard method to solve a large sparse linear equation system
is the conjugate-gradient method, which effectively generates a Krylov
space as does the Lanczos algorithm. The main effor in this method 
is to provide $\ham^2\impart \ket{c}$. Two remarks 
are in order. The reduced basis representation of $\ham^2$ is obtained by squaring the effective Hamiltonian 
generated by DMRG. This approximation is found to work extremely 
well as long as both real and imaginary part of the correction vector 
are included as target vectors: While the real part is not needed for 
the evaluation of spectral functions, $(E_{0}+\omega-\ham)\impart \ket{c} 
\sim \repart \ket{c}$ due to Eq.\ (\ref{eq:correctionvectorsystem}); and
targeting $\repart \ket{c}$ ensures minimal truncation errors in  
$\ham\impart \ket{c}$.
The fundamental drawback of using a 
squared Hamiltonian is that for all iterative eigenvalue or equation solvers
the speed of convergence is determined by the matrix condition number
which drastically deteriorates by the squaring of a matrix. This might
be avoided by using biconjugate or conjugate symmetric equation
solvers for Eq.\ (\ref{eq:sparseequation}) directly. 

Alternatively, there is a
reformulation of the correction vector method in terms of a 
minimization principle, which has been called
``dynamical DMRG'' \cite{Jeck02}. While the fundamental approach remains 
unchanged, the large sparse equation system is replaced by a 
minimization of the functional 
\begin{eqnarray}
    & & W_{A,\eta}(\omega,\psi) = \label{eq:minfunctional} \\
    & & \bra{\psi} (E_{0}+\omega-\ham)^2 + \eta^2 \ket{\psi} +
    \eta\langle A | \psi\rangle +
    \eta\langle \psi | A\rangle . \nonumber
\end{eqnarray}
At the minimum, the minimizing state is
\begin{equation}
    \ket{\psi_{{\rm min}}} = \impart \ket{c(\omega+\imag\eta)} .
\end{equation}
Even more importantly, the value of the functional itself is 
\begin{equation}
     W_{A,\eta}(\omega,\psi) = -\pi\eta C_{A} (\omega+\imag\eta) ,
\end{equation}
such that for the calculation of the spectral function it is not 
necessary to explicitly use the correction vector. In the simplest form of 
the correction vector method, the density matrix is 
formed from targeting four states, $\ket{0}$, 
$\hat{A}\ket{0}$, $\impart \ket{c(\omega+\imag\eta)}$ and
$\repart \ket{c(\omega+\imag\eta)}$.

\begin{figure}
    \includegraphics[angle=270,scale=0.32]{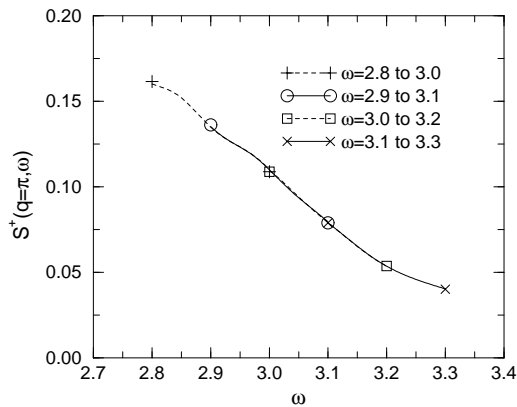}
\caption{Spectral weight of the $S=1/2$ Heisenberg 
AFM from correction vector DMRG. $M=256$ states kept. 
Spectral weights have been calculated for 
$\omega$-intervals starting from various anchoring frequencies for 
the correction vector. From \cite{Kuhn99}.}
\label{fig:correctionvector}
\end{figure}

As has been shown in \cite{Kuhn99}, it is not necessary to 
calculate a very dense set of correction vectors in $\omega$-space to 
obtain the spectral function for an entire frequency interval,
assuming that the finite 
convergence factor $\eta$ ensures that an entire range of energies of
width $\approx \eta$ is described quite well by the correction 
vector. It was found that best results are obtained for a two-correction vector 
approach where two correction vectors are calculated and targeted for 
two frequencies $\omega_{1}$, $\omega_{2}=\omega_{1}+\Delta \omega$ 
and the spectral function is obtained for the interval 
$[\omega_{1},\omega_{2}]$ using the Lanczos method for the approximate
Hamiltonian produced by this targeting scheme. 
This method is, for example, able to provide 
a high precision result for the spinon continuum in the $S=1/2$ 
Heisenberg chain where standard Lanczos dynamics fails (Fig. 
\ref{fig:singlemagnon_spin12}). 

\section{Early attempts at time-evolution}
Even though the methods described in the previous section provide
high-quality linear-response quantities, they fail in truly
out-of-equilibrium situations or for time-dependent Hamiltonians;
where they work, they are very time-consuming. It has therefore been
of high interest to find DMRG approaches dealing with state evolution 
in real-time. 

To see the advantages of such an approach, consider the following.
Essentially all physical quantities of interest involving time can be reduced 
to the calculation of either {\em equal-time} $n$-point correlators such as the
(1-point) density
\begin{equation}
  \langle n_i(t) \rangle = \bra{\psi(t)} n_i \ket{\psi(t)} = 
  \bra{\psi} e^{\imag\ham t} n_i e^{-\imag \ham t} \ket{\psi}
\label{eq:equaltime}
\end{equation} 
or {\em unequal-time} $n$-point correlators such as the (2-point) real-time
Green's function
\begin{equation}
    G_{ij} (t) = 
\bra{\psi} c^\dagger_{i}(t) c_{j}(0) \ket{\psi} = \bra{\psi} e^{+\imag\ham 
    t} c_{i}^\dagger e^{-\imag \ham t} c_{j} \ket{\psi} .
\label{eq:unequaltime}
\end{equation}    
This expression can be cast in a form very close to Eq.\ 
(\ref{eq:equaltime}) by introducing $\ket{\phi}=c_{j}\ket{\psi}$ such that
the desired correlator is then simply given as an equal-time matrix element
between two time-evolved states,
\begin{equation}
    G_{ij}(t) = \bra{\psi(t)} c_{i}^\dagger \ket{\phi(t)} .
\label{eq:matrixelement}
\end{equation}
If both $\ket{\psi(t)}$ and $\ket{\phi(t)}$ can be calculated, a very 
appealing feature of this approach is that $G_{ij}(t)$ can be evaluated
in {\em a single calculation} for all $i$ and $t$ as time proceeds. 
Frequency-momentum space is then reached by a double Fourier 
transformation. Obviously, finite system-sizes and edge effects as well
as algorithmic constraints will impose physical 
constraints on the largest times and distances $|i-j|$ or minimal 
frequency and wave vectors resolutions accessible. 
Nevertheless, this approach might
emerge as a very attractive alternative to the current very time-consuming
calculations of $G(k,\omega)$ using the dynamical DMRG\cite{Kuhn99,Jeck02}. 

The fundamental difficulty of obtaining the above correlators becomes
obvious if we examine the time-evolution of the quantum state 
$\ket{\psi(t=0)}$ under the action of some (for simplicity)
time-independent Hamiltonian $\ham\ket{\psi_{n}}=E_{n}\ket{\psi_{n}}$.
If the eigenstates $\ket{\psi_{n}}$ are known, expanding $\ket{\psi(t=0)}=
\sum_{n}c_{n}\ket{\psi_{n}}$ leads to the well-known time evolution
\begin{equation}
    \ket{\psi(t)}=\sum_{n}c_{n}\exp (-\imag E_{n}t) \ket{\psi_{n}} , 
\label{eq:simpletime}
\end{equation}
where the modulus of the expansion coefficients of $\ket{\psi(t)}$ is 
time-independent. A sensible Hilbert space truncation is given by a 
projection onto the large-modulus eigenstates. In strongly correlated 
systems, however, we usually have no good knowledge of the 
eigenstates. Instead, one uses some orthonormal basis with unknown 
eigenbasis expansion, $\ket{k}=\sum_{n} a_{kn} \ket{\psi_{n}}$. The 
time evolution of the state $\ket{\psi(t=0)}=
\sum_{k}d_{k}(0)\ket{k}$ then reads
\begin{equation}
    \ket{\psi(t)}=\sum_{k}\left( \sum_{n} d_{k}(0) a_{kn} e^{-\imag 
    E_{n}t} \right) \ket{k} \equiv \sum_{k} d_{k}(t) \ket{k} , 
\label{eq:complicatedtime}
\end{equation}
where the modulus of the expansion coefficients $d_k(t)$ is {\em 
time-dependent}. For a general orthonormal basis, 
Hilbert space truncation at one 
fixed time (i.e. $t=0$) will therefore not ensure a reliable 
approximation of the time evolution. Also, 
energy {\em differences} matter in time evolution due to the
phase factors $e^{-\imag (E_{n}-E_{n'})t}$ in $|d_{k}(t)|^2$. 
Thus, a good approximation to the low-energy Hamiltonian alone (as
provided by DMRG) is of limited use. 

{\em Static time-dependent DMRG.} 
Cazalilla and Marston\cite{Caza02} were the first to exploit DMRG to
systematically calculate time-dependent quantum many-body effects.
They studied a time-dependent Hamiltonian 
$\ham(t)\equiv \ham(0) + \hat{V}(t)$, where $\hat{V}(t)$ encodes the 
time-dependent part of the Hamiltonian.
After applying a standard DMRG calculation to the Hamiltonian 
$\ham(t=0)$, the time-dependent Schr\"{o}dinger equation was 
numerically integrated forward in time. The effective Hamiltonian
in the reduced Hilbert space was built as $\ham_{{\rm eff}}(t)=
\ham_{{\rm eff}}(0)+\hat{V}_{{\rm eff}}(t)$, where 
$\ham_{{\rm eff}}(0)$ was taken as the last superblock Hamiltonian 
approximating $\ham(0)$. $\hat{V}_{{\rm eff}}(t)$ as an approximation to 
$\hat{V}$ was built using the representations of operators in the final block 
bases. The initial condition was obviously to take 
$\ket{\psi(0)}$ as the ground state obtained by the preliminary DMRG run.
This procedure amounts to working within a {\em static} reduced Hilbert
space, namely that optimal
at $t=0$, and projecting all wave functions and operators onto it.

In this approach the hope is that an effective Hamiltonian 
obtained by targeting the ground state of the $t=0$ Hamiltonian is 
capable to catch the states that will be visited by the 
time-dependent Hamiltonian during time evolution. This approach
must however break down after relatively short times as the full Hilbert
space is explored, as became quickly obvious.

{\em Dynamic time-dependent DMRG.}
Several attempts have been made to improve on static time-dependent
DMRG by enlarging the reduced Hilbert space using
information on the time-evolution, such that the time-evolving state has
large support on that {\em dynamic} Hilbert space for longer times. Whatever procedure
for enlargement is used, the problem remains that the number of DMRG states
$m$ grows with the desired simulation time as they have to encode more and
more different physical states. As calculation time scales as $m^3$, this 
type of approach will meet its limitations somewhat later in time. 

All enlargement procedures rest on the ability of DMRG to describe -- at some
numerical expense -- small sets of states (``target states'') very well 
instead of just one. 

The simplest approach
is to target the set $\{\ket{\psi_i}\} = \{
\ket{\psi(0)}, \ham\ket{\psi(0)}, \ham^2\ket{\psi(0)}, \ldots\}$. 
Alternatively,
one might consider the Krylov vectors formed from this set. Results 
improve, but not decisively.

A much more time-consuming, but also much better performing 
approach has been demonstrated by Luo, Xiang and Wang 
\cite{Luo03}. They use 
a density matrix that is given by a superposition of states 
$\ket{\psi(t_{i})}$ at various times of the evolution,
$\dm = \sum_{i=0}^{N_{t}} \alpha_{i} 
\ket{\psi(t_{i})}\bra{\psi(t_{i})}$
with $\sum \alpha_{i}=1$
for the determination of the reduced Hilbert space.
Of course, these states    
are not known initially; it was proposed by them to
start within the framework of infinite-system DMRG from a small DMRG system and 
evolve it in time. For a very small system this procedure is exact.
For this system size, the state 
vectors $\ket{\psi(t_{i})}$ are used to form the density matrix. This
density matrix then determines the reduced 
Hilbert space for the next larger system, taking into account how 
time-evolution explores the Hilbert space for the smaller system.
One then moves on to the 
next larger DMRG system where the procedure is repeated. This is of course
very time-consuming. 

Schmitteckert\cite{Schm04} has computed the transport through a small 
interacting nanostructure using an Hilbert space enlarging approach,
based on the time evolution operator. To this end, 
he splits the problem into two parts: By obtaining a relatively large number of 
low-lying eigenstates exactly (within time-independent DMRG 
precision), one can calculate their time evolution exactly. 
For the subspace orthogonal to these eigenstates, he implements 
the matrix exponential $\ket{\psi(t+\Delta t)}=\exp (-\imag\ham \Delta 
t)\ket{\psi(t)}$ using the Krylov subspace approximation. For any block-site
configuration during sweeping, he evolves the state in time, obtaining
$\ket{\psi(t_{i})}$ at fixed times $t_{i}$. These are targeted in the
density matrix, such that upon sweeping forth and back a Hilbert space
suitable to describe all of them at good precision should be obtained.
For numerical efficiency, he carries out this procedure to convergence
for some small time, which is then increased upon sweeping, bringing
more and more states $\ket{\psi(t_{i})}$ into the density matrix.
Again, this is a very time-consuming approach.

\section{Time-evolving block decimation}

Decisive progress came from an unexpected corner, namely quantum
information theory, when Vidal proposed an algorithm for simulating 
quantum time evolutions of one-dimensional systems efficiently on a
classical computer \cite{Vida04,Vida03a}. His algorithm, known as TEBD [time-evolving
block decimation] algorithm, is based on matrix product
states\cite{Fann92,Klum93}; 
as it turned out, it is so closely linked to DMRG concepts, that his ideas
could be implemented easily into DMRG, leading to an {\em adaptive}
time-dependent DMRG, where the DMRG state space adapts itself in time 
to the time-evolving quantum state. In this section, we will explain
his algorithm.

A useful concept is that
of a {\em Schmidt decomposition:} Consider a quantum state $
\ket{\psi} = \sum_{ij} \psi_{ij} \ket{i} \otimes \ket{j}$ as
introduced before, with $N^{S}$ states $\ket{i}$ and $N^{E}$ states
$\ket{j}$. Assuming without loss of 
generality $N^S\geq N^E$, we form the $(N^S \times N^E)$-dimensional 
matrix $A$ with $A_{ij}=\psi_{ij}$. Singular value decomposition 
guarantees $A=UDV^T$, where $U$ is $(N^S \times N^E)$-dimensional with 
orthonormal columns, $D$ is a $(N^E \times N^E)$-dimensional diagonal 
matrix with non-negative entries $D_{\alpha\alpha}=\sqrt{w_{\alpha}}$, 
and $V^T$ is a $(N^E \times N^E)$-dimensional unitary matrix; $\ket{\psi}$
can be written as 
\begin{eqnarray}
    \ket{\psi} &=& 
    \sum_{i=1}^{N^S}\sum_{\alpha=1}^{N^E}\sum_{j=1}^{N^E} 
    U_{i\alpha} \sqrt{w_{\alpha}} V^T_{\alpha j} \ket{i}\ket{j} \\
    &=&
    \sum_{\alpha=1}^{N^E} \sqrt{w_{\alpha}} \left( \sum_{i=1}^{N^S} 
    U_{i\alpha} \ket{i} \right)
    \left( \sum_{j=1}^{N^E} V_{j\alpha} \ket{j} \right). \nonumber
\end{eqnarray}    
The orthonormality properties of $U$ and $V^T$ ensure that
$\ket{w_{\alpha}^S} = \sum_{i} U_{i\alpha}\ket{i}$ and 
$\ket{w_{\alpha}^E} = \sum_{j} V_{j\alpha}\ket{j}$ form orthonormal 
bases of system and environment respectively, in which the Schmidt 
decomposition
\begin{equation}
    \ket{\psi} = \sum_{\alpha=1}^{N_{{\rm Schmidt}}} \sqrt{w_{\alpha}}
    \ket{w_{\alpha}^S}\ket{w_{\alpha}^E} 
    \label{eq:Schmidtdecomp}
\end{equation}
holds. $N^{S}N^{E}$ coefficients $\psi_{ij}$ are reduced to 
$N_{{\rm Schmidt}}\leq N^E$ non-zero coefficients $\sqrt{w_{\alpha}}$,
$w_{1}\geq w_{2}\geq w_{3} \geq \ldots$. Relaxing the 
assumption $N^S\geq N^E$, one has
\begin{equation}
    N_{{\rm Schmidt}} \leq \min (N^{S},N^{E}) .
    \label{eq:Schmidtmax}
\end{equation}
Upon tracing out environment or system the reduced density matrices for 
system and environment are found to be
\begin{equation}
    \dm_{S} = \sum_{\alpha}^{N_{{\rm Schmidt}}} w_{\alpha} 
    \ket{w_{\alpha}^S}\bra{w_{\alpha}^S};
    \quad
    \dm_{E} = \sum_{\alpha}^{N_{{\rm Schmidt}}} w_{\alpha} 
    \ket{w_{\alpha}^E}\bra{w_{\alpha}^E}.
    \label{eq:dmbySchmidt}
\end{equation}
DMRG reduced density matrix analysis and the Schmidt decomposition
therefore yield exactly the same information. This fact was understood
from the very beginning of DMRG, although we had not heard the term 
``Schmidt decomposition''.
In fact,  the singular value decomposition representation of the wavefunction
was understood before the density matrix representation.

Let us now formulate the TEBD simulation algorithm. 
In the original exposition of the algorithm \cite{Vida03a}, 
one starts from a representation of a quantum state 
$\ket{\psi} = \sum_{\sigma_{1}\ldots\sigma_{L}} 
\psi_{\sigma_1,\ldots,\sigma_L} \ket{\sigma_{1}\ldots\sigma_{L}}$ where 
the coefficients for the states are decomposed as a product of tensors,
\begin{equation}
\psi_{\sigma_1,\ldots,\sigma_L}=\sum_{\alpha_1,\ldots,\alpha_{L-1}} 
\Gamma^{1}_{\alpha_1}[\sigma_1] \lambda^{1}_{\alpha_1} 
\Gamma^{2 }_{\alpha_1
\alpha_2}[\sigma_2]\lambda^{2}_{\alpha_2} \Gamma^{3 }_{\alpha_2 \alpha_3}
[\sigma_3]\cdots
\Gamma^{L }_{\alpha_{L-1}}[\sigma_L].\label{vidalcoef}
\end{equation}

It is of no immediate concern to us how the
$\Gamma$ and $\lambda$ tensors are constructed explicitly for a given 
physical situation. Let us assume that they have been determined such 
that they approximate the true wave function close to the optimum 
obtainable within the class of wave functions having such
coefficients; this is indeed possible as will be discussed below. 
There are, in fact, two ways of doing it: within the framework of 
DMRG, or by a continuous imaginary time 
evolution from some simple product state, as discussed in Ref.\ 
\cite{Vida04}. 

The ansatz can be visualized: the 
(diagonal) tensors $\lambda^{i}$, $i=1,\ldots,L-1$ are associated 
with the bonds $i$, whereas $\Gamma^{i}$, $i=2,\ldots,L-1$ links 
(transfers) from bond $i$ to bond $i-1$ across site $i$. Note that at 
the boundaries ($i=1,L$) the structure of the $\Gamma$ is modified. 
The sums run over $m$ states $\ket{\alpha_{i}}$ living in auxiliary state 
spaces on bond $i$. A priori, these states have no physical meaning.

\begin{figure}
        \includegraphics[scale=0.9]{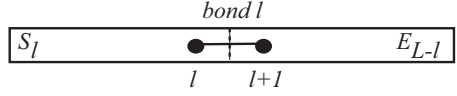}
\caption{Bipartitioning by cutting bond $l$ between sites $l$ and $l+1$.}
\label{fig:cut}
\end{figure}

The $\Gamma$ and $\lambda$ tensors are constructed such that for an
arbitrary cut of the system into a part $S_{l}$ of length $l$
and a part $E_{L-l}$ of length $L-l$ at bond $l$,
the Schmidt decomposition for this bipartite splitting reads
\begin{equation}
\ket{\psi}=\sum_{\alpha_l}\lambda^{l}_{\alpha_l}\ket{w_{\alpha_l}^{S_{l}}}
\ket{w_{\alpha_l}^{E_{L-l}}},
\end{equation}
with
\begin{equation}
\ket{w_{\alpha_l}^{S_{l}}}= \sum_{\alpha_1,\ldots,\alpha_{l-1}}
\sum_{\sigma_1,\ldots,\sigma_{l}} \Gamma^{1
}_{\alpha_1}[\sigma_1] \lambda^{1}_{\alpha_1} \cdots \Gamma^{l
}_{\alpha_{l-1}\alpha_l}[\sigma_l] 
\ket{\sigma_1}\otimes\cdots\otimes 
\ket{\sigma_l},
\label{eq:growsys}
\end{equation}
and
\begin{eqnarray}
\ket{w_{\alpha_l}^{E_{L-l}}}&=& \sum_{\alpha_l,\ldots,\alpha_{L-1}}
\sum_{\sigma_{l+1},\ldots,\sigma_{L}} \Gamma^{l+1
}_{\alpha_l \alpha_{l+1}}[\sigma_{l+1}]
\lambda^{l+1}_{\alpha_{l+1}}  \cdots \Gamma^{L
}_{\alpha_{L-1}}[\sigma_L] \times \nonumber \\
& & \ket{\sigma_{l+1}}\otimes\cdots\otimes
\ket{\sigma_L},\label{vidalphibdef}
\end{eqnarray}
where $\ket{\psi}$ is normalized and the sets of 
$\{\ket{w_{\alpha_l}^{S_{l}}} \}$ and $\{ \ket{w_{\alpha_l}^{E_{L-l}}}\}$
are orthonormal. This implies, for example, that
\begin{equation}
    \sum_{\alpha_{l}} (\lambda_{\alpha_{l}}^{l})^2 = 1.
\end{equation}    

We can see that (leaving aside normalization considerations for the 
moment, see \cite{Dale04}) 
this representation may be expressed as a matrix product state if we
choose for
$A^{i}[\sigma_i]=\sum_{\alpha,\beta}A^{i}_{\alpha\beta}[\sigma_{i}]\ket{\alpha}\bra{\beta}$
\begin{equation}
A_{\alpha\beta}^{i}[\sigma_i]=\Gamma^{i}_{\alpha\beta}[\sigma_i]
\lambda^{i}_\beta,
\end{equation}
except for $i=1$, and $i=L$, where expressions are slightly modified.

Let us now consider the time evolution for a typical (possibly time-dependent)
Hamiltonian with nearest-neighbor interactions:
\begin{equation}
\hat{H}=\sum_{i \: \rm odd}\hat{F}_{i,i+1} + \sum_{j \rm \: even}\hat{G}_{j,j+1},
\end{equation}
$\hat{F}_{i,i+1}$ and $\hat{G}_{j,j+1}$ are the local Hamiltonians on the odd bonds 
linking $i$ and $i+1$, and the even bonds linking $j$ and $j+1$. While all
$\hat F$ and $\hat G$ terms commute among each other, $\hat F$ and
$\hat G$ terms 
do in general not commute if they share one site.
Then the time evolution operator may be approximately 
represented by a (first order)
Trotter expansion as
\begin{equation}
\rmeh^{-\rmih \hat{H} \delta t}=\prod_{i \rm \: odd} \rmeh^{-\rmih \hat{F}_{i,i+1} \delta
t}\prod_{j \rm \: even} \rmeh^{-\rmih \hat{G}_{j,j+1} \delta t} + \mathcal{O}(\delta t^2),
\end{equation}
and the time evolution of the state can be computed by repeated application of the
two-site time evolution operators $\exp({-\rmih \hat{G}_{j,j+1} \delta t})$ and
$\exp({-\rmih \hat{F}_{i,i+1} \delta t})$. This is a well-known 
procedure in particular in Quantum Monte Carlo\cite{Suzu76} where it serves to 
carry out imaginary time evolutions (e.g. checkerboard decomposition). 

The TEBD simulation algorithm now runs as follows\cite{Vida04,Vida03a}:
\begin{enumerate}
    \item Perform the following two steps for all even bonds (the order does not
      matter):
\begin{itemize}
     \item[(i)] Apply $\exp({-\rmih \hat{G}_{l,l+1} \delta 
    t})$ to $\ket{\psi(t)}$. For each local time update, a new wave function is 
    obtained. The number of degrees of 
    freedom on the ``active'' bond thereby increases, as will be detailed below.  
    \item[(ii)] Carry out a Schmidt 
    decomposition cutting this bond and retain as in DMRG only 
    those $m$ degrees of freedom with the highest weight in the 
    decomposition. 
\end{itemize}
    \item Repeat this two-step procedure for all {\em odd} bonds, 
    applying $\exp({-\rmih \hat{F}_{l,l+1} \delta 
    t})$.
    \item This completes one Trotter time step. One may now evaluate 
    expectation values at selected time steps, and continues the 
    algorithm from step 1.
\end{enumerate}

Let us now consider some computational details.
\newline
(i) Consider a local time evolution operator acting on bond $l$, i.e.\ 
sites $l$ and $l+1$, for a state $\ket{\psi}$. The Schmidt 
decomposition of $\ket{\psi}$ after partitioning by cutting bond $l$ 
reads
\begin{equation}
    \ket{\psi}=\sum_{\alpha_{l}=1}^M \lambda^{l}_{\alpha_{l}}
    \ket{w_{\alpha_{l}}^{S_{l}}} \ket{w_{\alpha_{l}}^{E_{L-l}}} .
\label{oldstate}
\end{equation}
Using Eqs.\ (\ref{eq:growsys}) and (\ref{vidalphibdef}), we find after
expanding $\ket{w_{\alpha_{l}}^{S_{l}}}$ into $\ket{w_{\alpha_{l-1}}^{S_{l-1}}}$ 
and $\ket{\sigma_{l}}$, and similarly for $\ket{w_{\alpha_{l}}^{E_{L-l}}}$,  
\begin{eqnarray}
    \ket{\psi}&=& \sum_{\alpha_{l-1}\alpha_{l}\alpha_{l+1}} 
    \sum_{\sigma_{l}\sigma_{l+1}} 
    \lambda^{l-1}_{\alpha_{l-1}} 
    \Gamma^{l}_{\alpha_{l-1}\alpha_{l}}[\sigma_{l}] 
    \lambda^{l}_{\alpha_{l}} 
    \Gamma^{l+1}_{\alpha_{l}\alpha_{l+1}}[\sigma_{l+1}] 
    \lambda^{l+1}_{\alpha_{l+1}} 
    \times \nonumber \\
    & & \ket{w_{\alpha_{l-1}}^{S_{l-1}}} \ket{\sigma_{l}} \ket{\sigma_{l+1}}
    \ket{w_{\alpha_{l+1}}^{E_{L-(l+1)}}} .    
\end{eqnarray}    
We note, that this has the form of a typical DMRG state for two blocks and 
two sites
\begin{equation}
    \ket{\psi}=
    \sum_{m_{l-1}} 
    \sum_{\sigma_{l}} 
    \sum_{\sigma_{l+1}} 
    \sum_{m_{l+1}} \psi_{m_{l-1}\sigma_{l}\sigma_{l+1} m_{l+1}} 
    \ket{w^S_{m_{l-1}}}\ket{\sigma_{l}}\ket{\sigma_{l+1}}\ket{w^E_{m_{l+1}}} .
\end{equation}

The local time evolution operator on site $l, l+1$ can be expanded as  
\begin{equation}
    \hat{U}_{l,l+1} = \sum_{\sigma_{l}\sigma_{l+1}} 
    \sum_{\sigma_{l}'\sigma_{l+1}'} 
    U^{\sigma_{l}'\sigma_{l+1}'}_{\sigma_{l}\sigma_{l+1}}
    \ket{\sigma_{l}'\sigma_{l+1}'}\bra{\sigma_{l}\sigma_{l+1}}
\end{equation}
and generates $\ket{\psi'} = \hat{U}_{l,l+1} \ket{\psi}$. 
\begin{equation}
    \ket{\psi'}= \sum_{\alpha_{l-1}\alpha_{l+1}} 
    \sum_{\sigma_{l}\sigma_{l+1}} 
    \Theta^{\sigma_{l}\sigma_{l+1}}_{\alpha_{l-1}\alpha_{l+1}} 
    \ket{w_{\alpha_{l-1}}^{S_{l-1}}} \ket{\sigma_{l}} \ket{\sigma_{l+1}}
    \ket{w_{\alpha_{l+1}}^{E_{L-(l+1)}}} ,
\end{equation}    
where
\begin{equation}
    \Theta^{\sigma_{l}\sigma_{l+1}}_{\alpha_{l-1}\alpha_{l+1}}
=   \lambda^{l-1}_{\alpha_{l-1}}  \sum_{\alpha_{l}\sigma_{l}'\sigma_{l+1}'}
    \Gamma^{l}_{\alpha_{l-1}\alpha_{l}}[\sigma_{l}'] 
    \lambda^{l}_{\alpha_{l}}
        \Gamma^{l+1}_{\alpha_{l}\alpha_{l+1}}[\sigma_{l+1}'] 
	\lambda^{l+1}_{\alpha_{l+1}}
        U^{\sigma_{l}\sigma_{l+1}}_{\sigma_{l}'\sigma_{l+1}'} . 
\end{equation}
(ii) Now a {\em new} Schmidt decomposition identical to that in DMRG 
can be carried out for 
$\ket{\psi'}$: cutting once again bond $l$, there are now 
$mn_{{\rm site}}$ states in each part of the system, leading to 
\begin{equation}
    \ket{\psi'}=\sum_{\alpha_{l}=1}^{mn_{{\rm site}}} 
    \tilde{\lambda}^{l}_{\alpha_{l}}
    \ket{\tilde{w}_{\alpha_{l}}^{S_{l}}} 
    \ket{\tilde{w}_{\alpha_{l}}^{E_{L-l}}} .
\end{equation}
In general the states and 
coefficients of the decomposition will have changed compared to the 
decomposition (\ref{oldstate}) previous to the time evolution, and hence they are {\em 
adaptive}. We indicate this by introducing a tilde for these states 
and coefficients.
As in DMRG, if there are more than $m$ non-zero eigenvalues, we now choose the 
$m$ eigenvectors corresponding to the largest $\tilde{\lambda}^{l}_{\alpha_{l}}$ 
to use in these expressions. The error in the final state produced as a result 
is proportional to the sum of the magnitudes of the discarded 
eigenvalues. After normalization, to allow for the discarded 
weight, the state reads
\begin{equation}
    \ket{\psi'}=\sum_{\alpha_{l}=1}^{m} 
    \lambda^{l}_{\alpha_{l}}
    \ket{w_{\alpha_{l}}^{S_{l}}} 
    \ket{w_{\alpha_{l}}^{E_{L-l}}}.
\end{equation}
Note again that the states and coefficients in this superposition are in general different from 
those in Eq.\ (\ref{oldstate}); we have now dropped the tildes again, as 
this superposition will be the starting point for the next time 
evolution (state adaption) step.

The key point about the TEBD simulation algorithm is that a 
DMRG-style truncation to 
keep the most relevant density matrix eigenstates (or the maximum 
amount of entanglement) is carried out {\em at each time step.} 
This is in contrast to previous time-dependent DMRG methods, 
where the basis states were chosen before the time
evolution, and did not ``adapt'' to optimally represent the state at each
instant of time.

\section{Adaptive time-dependent DMRG}
\label{sec:newtime}

DMRG generates position-dependent $m\times m$ matrix-product states as block 
states for a reduced Hilbert space of $m$ states; the auxiliary state space 
to a bond is given by the Hilbert space of the block at whose end the bond sits. 
This physical meaning attached to the auxiliary state spaces implies 
that they carry good quantum numbers for all block sizes. The big 
advantage is that using good quantum numbers allows us to exclude a 
large amount of wave function coefficients as being 0, drastically speeding
up all calculations by at least one, and often two orders of magnitude. 

\begin{figure}
\includegraphics[scale=0.72]{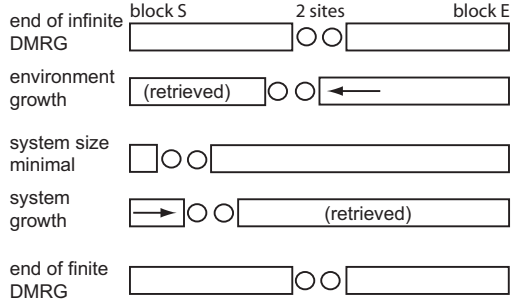}
\caption{Finite-system DMRG algorithm. Block growth and shrinkage. 
For the adaptive time-dependent DMRG, replace ground state 
optimization by local time evolution.}
\label{fig:finitesize}
\end{figure}

The effect of the finite-system DMRG algorithm\cite{Whit93} is now to shift the two 
free sites through the chain, growing and shrinking the blocks S and E as
illustrated in Fig. \ref{fig:finitesize}.
At each step, the ground state is redetermined and a new Schmidt 
decomposition carried out in which the system is cut between the two free sites, 
leading to a new truncation and new reduced basis transformations (2 matrices 
$A$ adjacent to this bond). 

As the actual decomposition and truncation 
procedure in DMRG and the TEBD simulation algorithm are identical, one
can use the finite-system algorithm to carry out the sequence of 
local time evolutions (instead of, or after, optimizing the ground state), 
thus constructing by Schmidt decomposition and 
truncation new block states best adapted to a state at any given 
point in the time evolution (hence 
adaptive block states) as in the TEBD algorithm, 
while maintaining the computational efficiency
of DMRG. To do this, one needs not only all reduced basis 
transformations, but also the wave function $\ket{\psi}$ in a two-block 
two-site configuration such that the bond that is currently updated 
consists of the two free sites. This implies that $\ket{\psi}$ has to 
be transformed between different configurations. In finite-system 
DMRG such a transformation, which was first implemented 
by White\cite{Whit96b} (``state 
prediction'') is routinely used to predict the outcome of large sparse 
matrix diagonalizations, which no longer occur during time 
evolution. Here, it merely serves as a basis transformation. 

The adaptive time-dependent DMRG algorithm which incorporates the 
TEBD simulation algorithm in the DMRG framework 
is therefore set up as follows:
\begin{itemize}
\item[0.] Set up a conventional finite-system DMRG algorithm with state 
prediction using the 
Hamiltonian at time $t=0$, $\ham(0)$, to determine the ground state of
some system of length $L$ using effective block Hilbert spaces of 
dimension $M$. At the end of this stage of the algorithm, we have
for blocks of all sizes $l$ reduced orthonormal 
bases spanned by states $\ket{m_{l}}$, which are characterized by good quantum 
numbers. Also, we have all reduced basis transformations, 
corresponding to the matrices $A$.
\item[1.]  For each Trotter time step, use the finite-system DMRG 
algorithm to run one sweep with the following modifications: 

 \begin{itemize}
  \item[i)] For each even bond apply the local time evolution $\hat{U}$ at the 
bond formed by the free sites to $\ket{\psi}$. This is a very fast operation
compared to determining the ground state, which is usually done instead in the
finite-system algorithm.
\item[ii)] As always, perform a DMRG truncation at each step of the 
finite-system algorithm, hence $O(L)$ times.
\item[(iii)] Use White's prediction method to shift the free sites by one.
\end{itemize}
\item[2.] In the reverse direction, apply step (i) to all odd bonds.
\item[3.] As in standard finite-system DMRG evaluate operators when 
desired at the end of some time steps. Note that there is no need to 
generate these operators at all those time steps where no operator 
evaluation is desired, which will, due to the small Trotter time step, 
be the overwhelming majority of steps. 
\end{itemize}    

Note that one can also perform every bond evolution operator at each half-sweep,
in order. This does not worsen the Trotter error, since in the reverse sweep
the operators are applied in reverse order.

The calculation time of adaptive time-dependent DMRG scales linearly in $L$,
as opposed to the static time-dependent DMRG which does not depend on 
$L$. The diagonalization of the density matrices (Schmidt 
decomposition) scales as $n_{{\rm site}}^3 m^3$; the preparation of 
the local time evolution operator as $n_{{\rm site}}^6$, but this may 
have to be done only rarely e.g.\ for discontinuous changes of 
interaction parameters. Carrying out the local time evolution
scales as $n_{{\rm site}}^4 m^2$; the basis transformation 
scales as $n_{{\rm site}}^2 m^3$. As $m\gg n_{{\rm site}}$ 
typically, the algorithm is of order $O(Ln_{{\rm site}}^3 m^3)$ at 
each time step.

The performance of this method has been tested in various applications
in the context of ultracold atom physics \cite{Dale04,Koll04,Koll05,Treb05}, 
but also for far-from-equilibrium dynamics \cite{Gobe04} and for
spectral functions \cite{Whit04}; some of these applications will
serve as examples in the following.
\section{Far-from-equilibrium dynamics}

In this section, we consider the dynamics of a system far from
equilbrium using adaptive time-dependent DMRG\cite{Gobe04}. The
following example, for which an exact solution is available, shows
that time-dependent DMRG can also perform in situations where
dynamical DMRG must surely fail.

The initial state $\ket{\init}=\ini$ 
on the one-dimensional spin-$1/2$ chains is subjected to 
the dynamics of the Heisenberg model
\be
\label{Hamilton}
H = \sum_n S^x_n S^x_{n+1} + S^y_n S^y_{n+1} + J_z S^z_n S^z_{n+1}
\equiv \sum_n h_n.
\ee
We set $\hbar = 1$, defining time to be 1/energy with
the energy unit chosen as the $J_{xy}$ interaction. 

Often it is useful to map the Heisenberg model
onto a model of interacting spinless fermions with nearest-neighbour
hopping:
\begin{eqnarray}
H&= &\sum_n  \left[ \frac{1}{2} ( c_n^\dagger c_{n+1}^\pdag +
  c_{n+1}^\dagger c_n^\pdag)\right. \nonumber \\
&&\left. + J_z
(c^\dagger_n c_n^\pdag- \frac{1}{2}) (c^\dagger_{n+1} c^\pdag_{n+1}- \frac{1}{2})\right]. 
\end{eqnarray}
In particular, the case $J_z=0$ describes free fermions on a lattice, and can 
be solved exactly. In the following we will focus on this case. Note
that in that case the initial state with two large ferromagnetic domains
separated by a domain wall in the center is a
highly excited state; the ground state exhibits power-law decaying
antiferromagnetic correlations.

The time evolution
delocalizes the domain wall over the entire chain; the
magnetization profile for the initial state $\ketini$ reads 
 \cite{Anta99}: 
\be
\label{Sz_nt}
S_z(n,t) = \bra{\psi(t)} S_n^z \ket{\psi(t)}
 =  -1/2\sum_{j=1-n}^{n-1}J_j^2(t),
\ee
where $J_j$ is the Bessel function of the first kind. 
$n=\ldots,-3,-2,-1,0,1,2,3,\ldots$ labels chain sites with the
convention that the first site in the right half of the chain has label 
$n=1$. As the total energy of the system is conserved,
the state cannot relax to the ground state. The exact solution reveals
a nontrivial behaviour with a complicated substructure in the magnetization
profile, which is a good benchmark for DMRG.

{\em Possible errors.}
Two main sources of error occur in the adaptive t-DMRG:
\\ (i) The {\em Trotter error} due to the Trotter decomposition. For
an $n$th-order Trotter decomposition
\cite{Suzu76}, the error made in one time step 
$dt$ is of order $L dt^{n+1}$. To reach a given time $t$ one has to perform 
$t/dt$ time-steps, such that in the worst case the error grows linearly in time $t$ and
the resulting error is of order $ L(dt)^n t$. 
\\
(ii) The DMRG {\em truncation error} due to the representation of the
time-evolving quantum state in reduced (albeit ``optimally'' chosen) 
Hilbert spaces and to the repeated transformations between different
truncated basis sets.
While the truncation error $\epsilon$ that sets the scale of
the error of the wave function and operators is typically very small, 
here it will strongly accumulate
as $O(Lt/dt)$ truncations are carried out up to time $t$. This is because
the truncated DMRG wave function has norm less than one and is renormalized
at each truncation by a factor of $(1-\epsilon)^{-1}>1$. Truncation errors
should therefore accumulate roughly exponentially with an exponent
of $\epsilon Lt/dt$, such that 
eventually the adaptive t-DMRG will break down at too long times.
The accumulated truncation error should decrease considerably with an 
increasing number of kept DMRG states $m$. For a fixed time $t$, 
it should decrease as the Trotter time step $dt$ is increased, as the number 
of truncations decreases with the number of time steps $t/dt$.

At this point, it is worthwhile to mention that our subsequent error analysis
should also be pertinent to the very closely related time-evolution
algorithm introduced by Verstraete {\em et al.}\cite{Vers04a}, which
also involves both Trotter and truncation errors.

We remind the reader that no error is encountered in the application of
the local time evolution operator $U_n$ to the state $\ket{\psi}$.

{\em Error analysis.} We use two main measures for the error: 
\\ (i) As a measure for the overall error we consider the {\em magnetization
deviation}, the maximum deviation of the local magnetization found by DMRG from
the exact result, 
\be
\mathrm{err}(t) = 
\mathrm{max}_n |\langle S^z_{n,\mathrm{DMRG}}(t)\rangle -
\langle S^z_{n,\mathrm{exact}}(t)\rangle |.
\ee
\\ (ii) As a measure which excludes the Trotter error we use the 
{\em forth-back deviation} $FB(t)$, 
which we define as the deviation between the initial state 
$\ketini$ and the state $\ket{fb(t)} = U(-t)U(t) \ketini$,
i.e.\ the state obtained by evolving $\ketini$ to some time $t$ and
then back to $t=0$ again. 
If we Trotter-decompose the time evolution operator $U(-t)$ into
odd and even bonds in the reverse order of the decomposition of $U(t)$,
the identity $U(-t) = U(t)^{-1}$ holds without any Trotter error, and
the forth-back deviation has the appealing property to capture the
truncation error only.

As the DMRG setup used in this particular calculation did not allow easy 
access to the fidelity
$|\bra{\init}fb(t) \rangle|$ (a calculation which is not a problem in
principle, see \cite{Treb05}), 
the forth-back deviation was defined
to be the $L_2$
measure for the difference of the magnetization profiles of
$\ketini$ and $\ket{fb(t)}$,
\be
FB(t) = \left( \sum_n \left( \bra{\mathrm{ini}} S^z_n \ketini - \bra{fb(t)}
S^z_n \ket{fb(t)} \right)^2 \right)^{1/2}
\!\!\!\!\!\!\!.
\ee

In order to control Trotter and truncation error, two DMRG control parameters
are available, the number of DMRG states $m$ and the Trotter time step $dt$.

\begin{figure}
\includegraphics[scale=0.5]{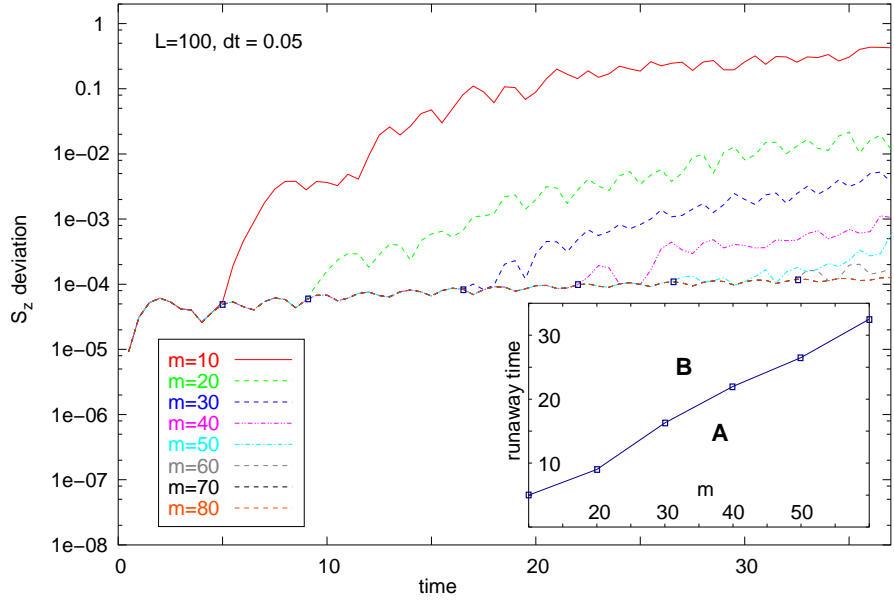}
 \caption{ 
Magnetization deviation ${\rm err}(t)$ as a function of time for
different numbers  
$m$ of DMRG states. The Trotter time interval is fixed at $dt = 0.05$. 
Again, two regimes can be distinguished: 
For early times, for which the Trotter error dominates, the error is slowly  
growing (essentially linearly) and independent of $m$ (regime A); 
for later times, the error is entirely given by the truncation error,
which is $m$-dependent and growing fast 
(almost exponential up to some saturation; regime B).
The transition between the two regimes occurs at a well-defined
``runaway time'' $t_R$ (small squares). The inset shows a monotonic, roughly
linear dependence of $t_R$ on $m$. From \cite{Gobe04}.}\label{eval_T50_m}
\end{figure}

The dependence on $dt$ is twofold: on the one hand, decreasing $dt$
reduces the Trotter error by some power of $dt^{n}$ exactly as in QMC; 
on the other
hand, the number of truncations increases, such that the truncation
error is enhanced. It is therefore not a good strategy to choose $dt$ 
as small as possible. The truncation error can however be decreased by
increasing $m$.

Consider the dependence of the magnetization deviation $\mathrm{err}(t)$
on the number $m$ of DMRG states.
In \Fig{eval_T50_m}, $\mathrm{err}(t)$ is plotted for a fixed Trotter time 
step $dt=0.05$ and different values of $m$. 
One sees that a $m$-dependent ``runaway time'' $t_R$ separates two
regimes: for $t<t_R$ (regime A), the deviation grows essentially linearly
in time and is independent of $m$, for $t>t_R$ (regime B), it suddenly
starts to grow more rapidly than any power-law as expected of the
truncation error.
In the inset of \Fig{eval_T50_m}, $t_R$ 
is seen to increase roughly linearly with growing $m$. As $m\rightarrow\infty$
corresponds to the complete absence of the truncation error, the
$m$-independent bottom
curve of \Fig{eval_T50_m} is a measure for the deviation due to the Trotter 
error alone and the runaway time can be read off very precisely as the moment
in time when the truncation error starts to dominate. 

\begin{figure}
\includegraphics[scale=0.5]{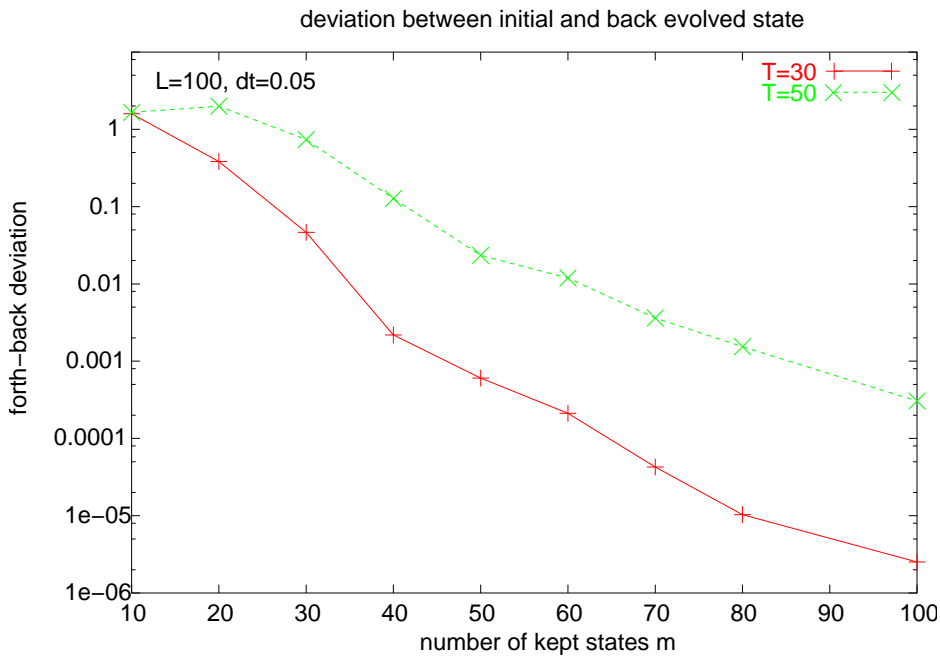}
\caption{ 
The forth-back error $FB(t)$ for $t=50$ and $t=30$, as function of
$m$.
Here, $L=100$, $dt = 0.05$. From \cite{Gobe04}.
}\label{backforth_m_fig}
\end{figure}

That the crossover from a dominating Trotter error at short times and a
dominating truncation error at long times is so sharp may seem surprising
at first, but can be explained easily by observing that the Trotter error
grows only linearly in time, but the accumulated truncation error grows 
almost exponentially in time.

To see that nothing special is happening at $t_{R}$, 
consider also \Fig{backforth_m_fig}, where the Trotter-error
free $FB(t)$ is plotted as 
a function of $m$, for $t=30$ and $t=50$. An approximately exponential increase 
of the accuracy of the method with growing $m$ is observed for a fixed time.
Our numerical results that indicate a roughly linear time-dependence of
$t_R$ on $m$ (inset of \Fig{eval_T50_m}) are the consequence of some
balancing of very fast growth of precision with $m$ and decay of precision
with $t$. 

The runaway time thus indicates an imminent
breakdown of the method and is a good, albeit very conservative measure of
available simulation times. We expect the above error analysis for the 
adaptive t-DMRG to be generic for other models. The truncation error
will remain also in approaches that dispose of the Trotter error; maximally
reachable simulation times should therefore be roughly the similar.
Even if for high precision calculation the Trotter error may dominate 
for a long time, in the long run it is always the truncation error
that causes the breakdown of the method at some point in time. 

\section{Finite temperature}

After the previous discussion on the difficulties of simulating the 
time-evolution of pure states in subsets of large Hilbert spaces it may seem 
that the 
time-evolution of mixed states (density matrices) is completely out of
reach. It is however easy to see that a thermal density matrix 
$\hat{\rho}_\beta \equiv \exp [-\beta \ham]$ can be constructed as a pure
state in an enlarged Hilbert space and that Hamiltonian dynamics
of the density matrix can be calculated considering just this pure state
(dissipative dynamics being more complicated).
In the DMRG context, this has first been
pointed out by Verstraete, Garcia-R\'{\i}poll and Cirac\cite{Vers04a} and
Zwolak and Vidal\cite{Zwol04}, 
using essentially 
information-theoretical language; it has also been used previously in pure
statistical physics language in e.g.\ high-temperature series 
expansions\cite{Buhl00}.

To this end, consider the completely
mixed state $\hat{\rho}_0 \equiv 1$. Let us assume that the dimension
of the local physical state space $\{\ket{\sigma_i}\}$ of a physical site 
is $n$.
Introduce now a local auxiliary state space $\{\ket{\tau_i}\}$ of the same
dimension $n$ on an auxiliary site. 
The local physical site is thus replaced by a rung of two sites, and 
a one-dimensional chain by a two-leg ladder of physical and auxiliary sites
on top and bottom rungs. Prepare now each rung $i$ in the Bell state
\begin{equation}
  \ket{\psi_0^i} = \frac{1}{\sqrt{n}} 
  \left[ \sum_{\sigma_i=\tau_i}^n |\sigma_i\tau_i\rangle \right] .
\label{eq:maxentstate}
\end{equation}
Other choices of $\ket{\psi_0^i}$ are equally feasible, as long as they
maintain in their product states maximal entanglement between
physical states $\ket{\sigma_i}$ and auxiliary states $\ket{\tau_i}$. 
Evaluating now the expectation value of some
local operator $\hat{O}_\sigma^i$ acting on the physical state space with
respect to $\ket{\psi_0^i}$, one finds
\[
\bra{\psi_0^i} \hat{O}_\sigma^i \ket{\psi_0^i} =
\sum_{\sigma_i=\tau_i} \sum_{\sigma_i'=\tau_i'} 
\frac{1}{n} \left[ \bra{\sigma_i\tau_i} \hat{O}_\sigma^i \otimes 1_\tau^i 
\ket{\sigma_i'\tau_i'} \right]. 
\]
The double sum collapses to
\[
\bra{\psi_0^i} \hat{O}_\sigma^i \ket{\psi_0^i} = 
\frac{1}{n} \sum_{\sigma_i=1}^n 
\bra{\sigma_i} \hat{O}_\sigma^i  
\ket{\sigma_i},
\]
and we see that the expectation value of $\hat{O}_\sigma^i$ with respect to
the pure state $\ket{\psi_0^i}$ living on the product of physical and 
auxiliary space is identical 
to the expectation 
value of $\hat{O}_\sigma^i$ with
respect to the completely mixed local physical state, or
\begin{equation}
\langle \hat{O}_\sigma^i \rangle = {\rm Tr}_\sigma \hat{\rho}_0^i 
\hat{O}_\sigma^i
\end{equation}
where
\begin{equation}
\hat{\rho}_0^i = {\rm Tr}_{\tau} \ket{\psi_0^i} \bra{\psi_0^i}.
\label{eq:rungpurificationi}
\end{equation}
This generalizes from rung to ladder using the density operator
\begin{equation}
\hat{\rho}_0 = {\rm Tr}_\mathbf{\tau} \ket{\psi_0} \bra{\psi_0},
\label{eq:rungpurification}
\end{equation}
where 
\begin{equation}
\ket{\psi_0}=\prod_{i=1}^L \ket{\psi_0^i} 
\end{equation}
is the product of all local Bell states, and the conversion from ficticious
pure state to physical mixed state is achieved by tracing out all auxiliary
degrees of freedom. 

At finite temperatures $\beta>0$ one uses
\[
\hat{\rho}_\beta = e^{-\beta\ham/2} \cdot 1 \cdot e^{-\beta\ham/2} = 
{\rm Tr}_\mathbf{\tau} e^{-\beta\ham/2} \ket{\psi_0} \bra{\psi_0} 
e^{-\beta\ham/2},
\]
where we have used Eq.\ (\ref{eq:rungpurification}) and the observation that 
the trace can be pulled out as it acts on the auxiliary space and 
$e^{-\beta\ham/2}$ on the physical space. Hence,
\begin{equation}
\hat{\rho}_\beta = {\rm Tr}_\mathbf{\tau} \ket{\psi_\beta} \bra{\psi_\beta},
\label{eq:rungpurification1}
\end{equation}
where $\ket{\psi_\beta}=e^{-\beta\ham/2} \ket{\psi_0}$. Similarly, this
finite-temperature density matrix can now be evolved in time by
considering $\ket{\psi_\beta (t)}=e^{-i\ham t} \ket{\psi_\beta (0)}$ and
$\hat{\rho}_\beta(t) = {\rm Tr}_\mathbf{\tau} \ket{\psi_\beta(t)} 
\bra{\psi_\beta(t)}$. The calculation of the finite-temperature time-dependent
properties of, say, a Hubbard chain, therefore corresponds to the 
imaginary-time and real-time evolution of a Hubbard ladder prepared to be
in a product of special rung states. Time evolutions generated by Hamiltonians 
act on the physical 
leg of the ladder only. As for the evaluation of expectation values
both local and auxiliary degrees of freedom are traced on the same footing, 
the distinction can be completely dropped but for the time-evolution itself.
Code-reusage is thus almost trivial. 
Note also that the initial 
infinite-temperature pure state needs only $m=1$ block states to be described
exactly in DMRG as it is a product state of single local states. Imaginary-time
evolution (lowering the temperature) will introduce entanglement such that to
maintain some desired DMRG precision $m$ will have to be increased.

\section{Time-step targetting}
The Trotter based methods for time evolution discussed above, while very fast, 
have 
two notable weaknesses: first, there is an error proportional to the time step $\tau$
squared. This error is usually tolerable and can be reduced to neglible levels 
by using higher order
Trotter decompositions\cite{Whit04c}.  More importantly,
they are limited to systems with nearest neighbor
interactions on a single chain.  This limitation is more difficult to deal with. 
In the case of narrow ladders with
nearest-neighbor interactions, one
can avoid the problem by lumping all sites in a rung into a single supersite.
Another approach would be to use a superblock configuration with, say, three
center sites, which would allow one to treat two-leg ladders without using
supersites.
Unfortunately, these approaches become very inefficient for wider ladders, and are
not applicable at all to general long-range interaction terms.

The time-step targetted (TST) method does not have these limitations. The main idea is
to produce a basis
which targets the states needed to represent one small but finite time step. 
Once this basis
is complete enough, the time step is taken and the algorithm proceeds to the
next time step. This targetting is intermediate to previous approaches:
the Trotter methods target precisely one instant in time at any DMRG step,
while Luo, Xiang, and Wang's approach\cite{Luo03} targetted the entire
range of time to be studied. Targetting a wider range of time requires more
density matrix eigenstates be kept, slowing the calculation.
By targetting only a small interval of time, a smaller price is paid relative
to the most efficient Trotter methods. In exchange for
the modest loss of efficiency, we gain the ability to treat longer range
interactions, ladder systems, and narrow two-dimensional strips. In addition,
the error from a finite time step is greatly reduced relative to the second
order Trotter method.

The procedure of Luo, et. al. for targetting an interval of time is nearly ideal:
one divides the interval into $n$ small steps of length $\epsilon$, and targets
$\psi (t=0)$, $\psi (t=\epsilon )$, $\psi (t=2\epsilon )$, $\ldots$, $\psi(t=n\epsilon)$,
simultaneously. By targetting these wavefunctions simultaneously, any linear
combination of them is also included in the basis. This means than the basis
is able describe an $n+1$-th order interpolation through these points, making
it for reasonable $\epsilon$ and $n$ essentially complete over the time interval.
In the TST method the interval is short and $n$ is fairly small:
in the implementation of \cite{Whit04c}, $n=3$ and the time step is similar in
size to the Trotter step $\tau$, say $\sim J/10$ for a spin chain.

The Runge-Kutta (R-K) implementation of this approach is defined as follows:
one takes a tentative time step at each DMRG step, the
purpose of which is to generate a good basis. The standard fourth order
R-K algorithm is used. This is defined
in terms of a set of four vectors:
\begin{eqnarray}
|k_1\rangle &=& \tau  \tilde{H}(t) |\psi(t)\rangle, \nonumber \\
|k_2\rangle &=& \tau  \tilde{H}(t+\tau/2) \left[ |\psi(t)\rangle + 1/2 |k_1\rangle \right], 
\nonumber \\
|k_3\rangle &=& \tau  \tilde{H}(t+\tau/2) \left[ |\psi(t)\rangle + 1/2 |k_2\rangle \right], 
\nonumber \\
|k_4\rangle &=& \tau  \tilde{H}(t+\tau) \left[ |\psi(t)\rangle + |k_3\rangle \right],
\label{kvectors}
\end{eqnarray}
where $\tilde{H}(t) = H(t)-E_0$.
The state at time $t+\tau$ is given by
\begin{equation}
|\psi(t+\tau)\rangle \approx \frac{1}{6}
\left[ |k_1\rangle + 2|k_2\rangle + 2|k_3\rangle + |k_4\rangle \right] + O(\tau^5).
\end{equation}
\label{newpsi}
We target the state at times
$t$, $t+\tau/3$, $t+2\tau/3$ and $t+\tau$.
The R-K vectors have been chosen to minimize the error in $|\psi(t+\tau)\rangle$, but
they can also be used to generate $|\psi\rangle$ at other times.
The states at times $t+\tau/3$ and $t+2\tau/3$ can be approximated, with an
error $O(\tau^4)$, as
\begin{eqnarray}
|\psi(t+\tau/3)\rangle & \approx & |\psi(t)\rangle + \nonumber \\
& + & \frac{1}{162} \left[ 31|k_1\rangle +14 |k_2\rangle + 14 |k_3\rangle  - 5 |k_4\rangle 
\right], \nonumber \\
|\psi(t+2\tau/3)\rangle & \approx & |\psi(t)\rangle + \nonumber \\
& + & \frac{1}{81} \left[ 16|k_1\rangle + 20 |k_2\rangle + 20 |k_3\rangle  - 
2 |k_4\rangle \right].
\label{targets}
\end{eqnarray}

Each half-sweep corresponds to one time step.
At each step of the half-sweep, one calculates the R-K vectors
(\ref{kvectors}), but without advancing in time. The density matrix is then
obtained with the target states
$|\psi(t)\rangle$, $|\psi(t+\tau/3)\rangle$, $|\psi(t+2\tau/3)\rangle$, and
$|\psi(t+\tau)\rangle$.
Advancing in time is done on the last step of a half-sweep. However,
we may choose to advance in time only every other half-sweep, or only
after several half-sweeps, in order to make sure the basis adequately
represents the time-step. For the systems of Ref. \cite{Whit04c}, one half-sweep
was adequate and the most efficient.
The method used to advance in time in the last step need not be the R-K
method used in the previous tentative steps. In fact, the computation time involved
in the last step of a sweep is typically miniscule, so a more accurate
procedure is warranted.
A simple way to do this 
is to perform, say,  10 R-K iterations with step $\tau/10$. 
The relative weights of the states targetted can be optimized. An equal weighting
is not optimal; the initial time and final time are more important. In
Ref. \cite{Whit04c}, it was found that giving a weight of $1/3$ for the first
and final states, and $1/6$ for the two intermediate states, gave excellent results.

\begin{figure}
\includegraphics[scale=0.3]{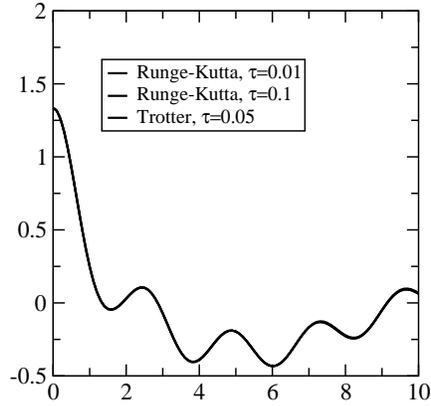}
\caption{ The value of $\bra S^-(16,t)S^+(16,0)\rangle$ computed for a 31 site 
$S=1$ Heisenberg chain,
computed three different times. Here the curves labeled Runge-Kutta are the 
TST method, implemented using Runge Kutta. The time step is $\tau$. 
The difference in results are not visible
on this scale.  }\label{compone}
\end{figure}

\begin{figure}
\includegraphics[scale=0.3]{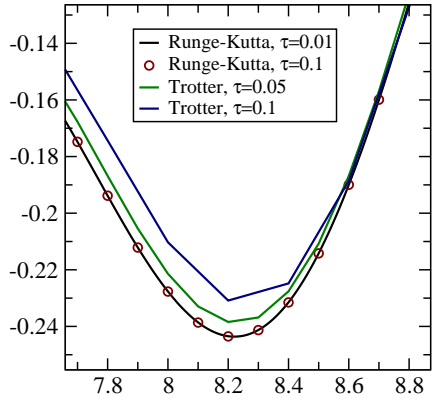}
\caption{ 
Same as for \Fig{compone}, but showing only a small region so the differences
become apparent.  }\label{comptwo}
\end{figure}

Both the Trotter method and the TST method give very accurate
results. In \Fig{compone} and \Fig{comptwo}, we show a comparison of the methods.
On a large scale, we cannot see any difference between the methods for times
out to $t \sim 10$. If we zoom in on a particular region, we see the effects
of the finite Trotter decomposition error, here falling as $\tau^2$. We kept
$m=300$ states for the TST method, and $m=200$ states for the Trotter methods. 
Typically, one finds that more states must be targetted for the TST method, because
the targetting is over a finite interval of time rather than one instant.
The Trotter decomposition error can be eliminated almost completely by using
a higher order decomposition. In this case, the smaller value of $m$ still works
as well as in the lower order methods. This combination gives the best combination
of speed and accuracy.

\section{Spectral functions}
Using either the Trotter or TST methods, it is straightforward
to obtain spectral functions.
Typically, we are interested in the Fourier transform of a time
dependent correlation function
\begin{equation}
C(t) = \bra{\phi}B(t) A(0)\ket{\phi}
\label{greens}
\end{equation}
where $\ket{\phi}$ is the ground state. 

It is convenient to write this (cf.\ Eq.\ (\ref{eq:unequaltime})) as
\begin{equation}
C(t) = \bra{\phi}B\exp[-i t (H-E_G)] A\ket{\phi}
\label{greentwo}
\end{equation}
where $E_G$ is the ground state energy. For evaluating this
expression, we proceed as follows: we first obtain
the ground state using the standard DMRG algorithm. We then
apply the operator $A$ to obtain $\ket{\psi(t=0)}$, and evolve $\ket{\psi(t)}$ in time,
using the Hamiltonian with the ground state energy subtracted off. 
During this time evolution, we target both $\ket{\psi(t)}$ and $\ket{\phi}$.
We then obtain $C(t)$, at each time step, as
\begin{equation}
C(t) = \bra{\phi}B \ket{\psi(t)}.
\label{greenthree}
\end{equation}
By targetting both $\ket{\psi(t)}$ and $\ket{\phi}$, we ensure that this
matrix element can be obtained with an accuracy controlled by the truncation
error.

In forming $\ket{\psi(t=0)}$, we use a complete half-sweep to apply $A$ to $\ket{\phi}$.
In particular, if $A$ is a sum of
terms $A_j$ over a number of sites, then we apply an $A_j$ only when $j$
is one of the two central, untruncated sites. Thus the basis is automatically
suitable for $A_j \ket{\phi}$.  During this buildup of $A$ at step $j$ we
target both
$\ket{\phi}$ and $\sum_{j'=1}^j A_{j'}\ket{\phi}$.
At the end
of the sweep, we turn on the time evolution. 

For translationally invariant systems
it is particularly convenient to let $A$ and $B$ be on-site operators, for example
$A = S^+(j)$,
where $j$ is in the center of the chain.  Since the time evolution does not
evolve $B$, a whole set of $B$'s can be utilized, one for each site of the system,
for example $B = S^-(\ell)$.
One measurement of $G(\ell-j,t)$ can be made on each step of each sweep, where $\ell$
is one of the two center sites with untruncated bases, so that no extra operator
matrices need be kept to reproduce $B$. In this way, one simulation yields 
$G(\ell-j,t)$ for a wide range of values of $\ell-j$ and $t$. By Fourier transforming
in both space and time, one obtains the full spectral function for all frequency
and momenta, in one simulation. This is in stark contrast to the most accurate frequency
methods, in which one $k$ and a small range of $\omega$ are obtained in one run. 

\begin{figure}
\includegraphics[scale=0.3]{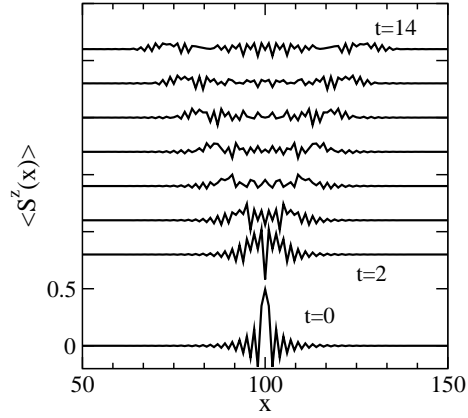}
\caption{ Time evolution of the local magnetization $\langle S^z(x) \rangle$
of a 200 site spin-1 Heisenberg chain after $S^+(100)$ is applied. From \cite{Whit04}.
}
\label{figspreadout}
\end{figure}

As an example we return to the isotropic spin-1 Heisenberg chain, with
the exchange coupling $J$ set to unity, and
$A = S^+(j)$, as above.
Note that the application of $S^+(j)$ constructs
a localized wavepacket consisting of all wavevectors. This packet spreads out
as time progresses, with different components moving at different speeds.
The speed of a component is its group velocity, determined as the slope
of the dispersion curve at $k$.
In Fig.\ref{figspreadout} we show the local magnetization
$\bra{\psi(t)}S^z\ket{\psi(t)}$ for a chain of length $L=200$, with timestep $\tau=0.1$.
At $t=0$, the wavepacket has
a finite extent, with size given by the spin-spin correlation length $\xi$.
At later times, the different speeds of the different components give
the irregular oscillations in the center of the packet.
We kept $m=150$ states per block, giving a truncation error of about
$6\times10^{-6}$. When the wavefront reaches the edge of the system, we
stop the simulation. Our results up to that point in time have very minimal
finite size effects, dying off exponentially from the edges. The correlation
function is exponentially small for $|\ell-j|$ greater than $v t$, with
$v$ the maximum group velocity. Because of this, we can specify the
momentum precisely and arbitrarily, i.e. as if the system were infinite.
The broadening from having a finite system appears only in frequency, not momentum. 

The spectral function is defined as $-1/\pi {\rm Im} G(k,\omega)$, where
\begin{equation}
G(x,t) \equiv -i C(x,t) \equiv  -i \bra{\phi} T[S^-_x(t) S^+_0(0)] \ket{\phi}
\end{equation}
Since $G(x,t)$ is even in $x$ and $t$, the Fourier transform is
\begin{equation}
G(k,\omega) = 2 \int_0^\infty dt \cos \omega t \sum_x \cos kx G(x,t)
\end{equation}
In this expression, it is the real part of $C(x,t)$
that determines the spectral function; the imaginary part is thrown away. For
an excited state with energy $\Delta$ above the ground state, this spectral
function gives peaks at $\pm \Delta$. Alternatively, we can define the
spectral function to have only the $+\Delta$ peaks. This spectral function
comes from a Fourier transform utilizing both the real and imaginary parts
of $C(x,t)$, and both positive and negative times:
\begin{equation}
A(k,\omega) = \frac{1}{2\pi}\int_{-\infty}^\infty dt e^{i \omega t} \sum_x\cos kx
\bra{\phi} S^-_x(t) S^+_0\ket{\phi}.
\end{equation}
In this case, we utilize $C(x,t) = C(x,-t)^*$ to obtain the negative time data.
We prefer this latter spectral function, since it utilizes the imaginary data,
but the differences are not large and we have not studied them carefully.

We approximate the time integral utilizing a windowing function $W(t)$
which goes to zero as $t \to T$,
\begin{equation}
\int_{-\infty}^\infty \approx \int_{-T}^T W(t).
\end{equation}
A set of windowing functions with a number of nice properties is
\begin{equation}
W_n(t) = \cos(\frac{\pi t}{2 T})^n.
\end{equation}
These functions approach Gaussians as $n\to\infty$, but the function and
$n-1$ derivatives vanish at $t=\pm T$. If one sets $W_n(t)$ to zero for $|t|>T$,
and Fourier transforms, one obtains a nearly Gaussian lineshape with oscillating
tails falling off as $\omega^{-(n+1)}$.  We have used $n=4$, for which the 
lineshape in $\omega$
has negative regions in the tails of very small amplitude, less than half
a percent of the peak height. Another
good choice is $n=3$, giving a somewhat narrower peak at the expense of 
more negativity.
Note that if the true spectral function has an isolated delta function peak,
the windowed spectrum will have a broadened peak centered
precisely at the same frequency.  Thus it is possible to locate the
single magnon line with an accuracy much better than $1/T$.
If a continuum is also present nearby, the peak is less well determined.
In the case of the $S=1$ chain, for $k$ near $\pi$ the peak is isolated,
but at some point near $k=0.25 \pi$ the peak enters the two magnon continuum
and develops a finite width.

\begin{figure}
\includegraphics[scale=0.3]{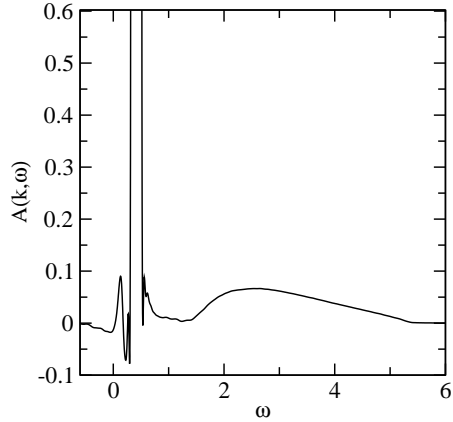}
\caption{
The single magnon spectrum of the spin-1 Heisenberg antiferromagnetic chain,
for a system of $L=600$ sites, using the Trotter method with 
a time step of $\tau=0.1$, running for $T=100$,
and keeping up to $m=600$ states, at
momentum $k=\pi$. 
The main peak has a height of 83 at $\omega=0.415$,
close to the true Haldane gap of $\Delta = 0.41050(2)$.\cite{whitehuse} 
The sharp oscillations
around it are the result of numerical errors and windowing.
The three-magnon continuum is visible, beginning at $3 \Delta$.\cite{whitehuse}
}
\label{figkpi}
\end{figure}

\begin{figure}
\includegraphics[scale=0.3]{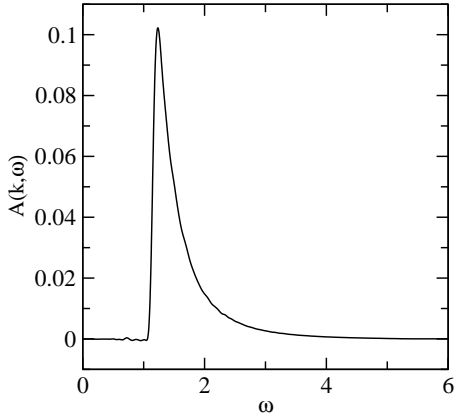}
\caption{
Same as for \Fig{figkpi}, but at $k=\pi/10$
}
\label{figktenth}
\end{figure}

In Fig.\ref{figkpi} we show the resulting spectrum for $k=\pi$. The results
for the three magnon continuum are impressive, as the single magnon line is
much larger in amplitude.
In Fig.\ref{figktenth} we show results for $k=\pi/10$. For this momentum,
the single magnon line lies within the two magnon continuum, altering its
shape dramatically. This shape has been calculated approximately using
the nonlinear sigma model\cite{weston}; the results shown are in good 
qualitatively agreement with these analytic results.

\section{Conclusion}

While the invention of efficient time-dependent DMRG methods is at the
time of writing only one and a half years old, the results achieved so
far have already been impressive, indicating that the problem of
highly precise time-evolutions for one-dimensional strongly correlated
quantum is for the first time under very good control. 
The available variants cover a
wide range of physical problems, with Trotter-based methods most efficient 
for short-ranged Hamiltonians, and with time-step targetting methods superior
for longer-ranged interactions. They also provide powerful
alternatives for the calculation of spectral functions in the
momentum-frequency range. A nice feature is provided by their easy
implementation in the framework of existing static finite-system DMRG 
codes. As control of quantum systems is improving
experimentally, we expect the range of physical applications to grow
strongly in the very near future. 

\begin{theacknowledgments}
SRW acknowledges the support of the NSF under grant DMR03-11843.
\end{theacknowledgments}


\begin{thebibliography}{99} 
\bibitem{Grei02} M. Greiner, O. Mandel, T. Esslinger, T. W. H\"{a}nsch and
I. Bloch, Nature (London) \textbf{415}, 39 (2002)

\bibitem{Kohl05} M. K\"{o}hl, H. Moritz, T. St\"{o}ferle, 
K. G\"{u}nter and T. Esslinger,
Phys. Rev. Lett. \textbf{94}, 080403 (2005)
    
\bibitem{Whit92} S.R. White, Phys.\ Rev.\ Lett.\ \textbf{69}, 2863
(1992)

\bibitem{Whit93} S.R. White, Phys.\ Rev.\ B \textbf{48}, 10345
(1993)

\bibitem{Scho04} U. Schollw\"{o}ck, Rev.\ Mod.\ Phys. \textbf{77}, 259 (2005)

\bibitem{Vida04} G. Vidal, Phys.\ Rev.\ Lett.\ \textbf{93}, 040502
(2004)

\bibitem{Dale04} A. J. Daley, C. Kollath, U. Schollw\"{o}ck and
G. Vidal, J. Stat.\ Mech.: Theor.\ Exp.\ (2004) P04005

\bibitem{Whit04} S. R. White and A. Feiguin, Phys.\ Rev.\ Lett.\
\textbf{93}, 076401 (2004)

\bibitem{Whit04c} A. Feiguin and S. R. White, Phys.\ Rev.\ B \textbf{72}, 
020404 (2005).

\bibitem{Wils75} K. G. Wilson, Rev.\ Mod.\ Phys.\  \textbf{47},
773 (1975) 

\bibitem{Ostl95} S. \"{O}stlund and S. Rommer, Phys.\ Rev.\ Lett.\ 
\textbf{75}, 3537 (1995)

\bibitem{Vers04} F. Verstraete, D. Porras and J. I. Cirac, 
Phys.\ Rev.\ Lett.\ \textbf{93}, 227205 (2004)

\bibitem{Hall95} K. Hallberg, Phys.\ Rev.\ B \textbf{52}, 9827
(1995)

\bibitem{Gagl87} E. R. Gagliano and C. A. Balseiro, Phys.\ Rev.\ Lett. 
\textbf{59}, 2999 (1987)

\bibitem{Rama97} S. Ramasesha, S. K. Pati, H. R. Krishnamurthy, Z.
Shuai, and J. L. Br\'{e}das, Synth. Met. \textbf{85}, 1019 (1997)

\bibitem{Kuhn99} T. K\"uhner and S.R. White, Phys.\ Rev.\ B
\textbf{60}, 335 (1999) 

\bibitem{Jeck02} E. Jeckelmann, Phys.\ Rev.\ B \textbf{66}, 045114
(2002)

\bibitem{Soos89} Z. G. Soos and S. Ramasesha, J. Chem.\ Phys.\ \textbf{90}, 
1067 (1989)

\bibitem{Caza02} M. Cazalilla and B. Marston, 
Phys.\ Rev.\ Lett.\ \textbf{88}, 256403 (2002) 

\bibitem{Luo03} H. G. Luo, T. Xiang and X. Q. Wang, Phys.\ Rev.\ Lett.\ 
\textbf{91}, 049701 (2003) 

\bibitem{Schm04} P. Schmitteckert, Phys.\ Rev.\ B \textbf{70}, 121302 
(2004)

\bibitem{Vida03a} G. Vidal, Phys.\ Rev.\ Lett. \textbf{91}, 147902
(2003)

\bibitem{Fann92} M. Fannes, B. Nachtergaele and R. F. Werner,   
Comm.\ Math.\ Phys.\ \textbf{144}, 3 (1992) 

\bibitem{Klum93} A. Kl\"{u}mper and A. Schadschneider and J. Zittartz,
Europhys.\ Lett.\ \textbf{24}, 293 (1993)   

\bibitem{Suzu76} M. Suzuki, Prog.\ Theor.\ Phys.\ \textbf{56}, 1454
(1976)

\bibitem{Whit96b} S. R. White, Phys.\ Rev.\ Lett.\ \textbf{77}, 3633
(1996)

\bibitem{Gobe04} D. Gobert, C. Kollath, U. Schollw\"{o}ck and G. Sch\"{u}tz,
Phys.\ Rev.\ E \textbf{71}, 036102 (2005) 

\bibitem{Anta99} T. Antal, Z. Racz, A. Rakos, G. Sch\"{u}tz,
Phys.\ Rev.\ E \textbf{59}, 4912 (1999)

\bibitem{Vers04a} F. Verstraete, J. J. Garcia-R\'{\i}poll and J. I. Cirac,
Phys.\ Rev.\ Lett.\ \textbf{93}, 207204 (2004)

\bibitem{Koll04} C. Kollath, U. Schollw\"{o}ck, J. von Delft and W. Zwerger,
Phys.\ Rev.\ A \textbf{71}, 053606 (2005) 

\bibitem{Koll05} C. Kollath, U. Schollw\"{o}ck and W. Zwerger,
Phys.\ Rev.\ Lett.\ \textbf{95}, 176401 (2005) 

\bibitem{Treb05} S. Trebst, U. Schollw\"{o}ck, M. Troyer and P.
Zoller, cond-mat/0506809

\bibitem{Zwol04} M. Zwolak and G. Vidal, 
Phys.\ Rev.\ Lett.\ \textbf{93}, 207205 (2004)

\bibitem{Buhl00} A. B\"{u}hler, N. Elstner and G. S. Uhrig,
Eur.\ Phys.\ J. B \textbf{16}, 475 (2000)

\bibitem{whitehuse} S.R. White and D.A. Huse, Phys.\ Rev.\ B \textbf{48},  
3844 (1993).

\bibitem{weston} I. Affleck and R.A. Weston, Phys.\ Rev.\ B \textbf{45},  
4667 (1992).


\end{thebibliography}
\end{document}